 \documentclass[12pt]{article}
\usepackage[dvips]{graphicx}

\catcode`@=11
\def\secteqno{\@addtoreset{equation}{section}%
\def\theequation{\thesection.\arabic{equation}}}
\catcode`@=12 \secteqno
\newcommand{\be}{\begin{equation}}
\newcommand{\ee}{\end{equation}}
\newcommand{\bea}{\begin{eqnarray}}
\newcommand{\eea}{\end{eqnarray}}
\newcommand{\bref}[1]{(\ref{#1})}
\newcommand{\nn}{\nonumber}

\newcommand{\bi}{\begin{enumerate}}
\newcommand{\ei}{\end{enumerate}}
\newcommand{\A}{\alpha}  
 \newcommand{\D}{\delta} 
\newcommand{\ep}{\epsilon} 
\newcommand{\T}{\theta} 
   
\newcommand{\lam}{\lambda}      
          \newcommand{\w}{\omega}
          
\newcommand{\h}{\eta}           
           
\newcommand{\W}{\Omega}

\def\kx{{\kappa}}\def\bx{{\beta}}

\def\6{\partial}\def\7{\tilde}\def\8{\hat}
\def\too{\quad\to\quad}
\def\pa{\partial}

\def\CL{{\cal L}}
\def\CH{{\cal H}}
\def\CP{{\cal P}}\def\CZ{{\cal Z}}\def\CJ{{\cal J}}\def\CO{{\cal O}}
\def\t{\tilde}\def\so{{so}}
\def\l{{\ell}}
\def\vs{\vskip 4mm}\def\={{\;=\;}}\def\+{{\;+\;}}
\newcommand{\Iso}{I{\hskip-0.05cm{so}}}

\def\CZ{{\cal Z}}
\def\CM{{\cal M}}

\def\Deq21{{$D \hskip-1mm = \hskip-1mm 2 \hskip-1mm + \hskip-1mm1$}}
\def\Dneq21{{$D \hskip-1mm\neq\hskip-1mm 2 \hskip-1mm + \hskip-1mm1$}}
\def\vs{\vskip 4mm}

\begin{document}

\hfill  ICCUB-09-210
UB-ECM-PF09/16 ,Toho-CP-0990

\vskip 20mm
\begin{center}
{\Large\bf Deformations of Maxwell algebra and their Dynamical
Realizations } \vskip 10mm {\large ~ Joaquim Gomis$^{1}$,  Kiyoshi
Kamimura$^2$ and Jerzy Lukierski$^3$}\vskip 6mm
\medskip

$^1$Departament d'Estructura i Constituents de la
Mat\`eria and ICCUB, Universitat de Barcelona, Diagonal 647, 08028 Barcelona\\
\vspace{6pt} $^2$  Department of Physics, Toho University,
Funabashi, 274-8510 Japan\\
\vspace{6pt}$^3$Institute of Theoretical
 Physics, Wroclaw University, pl. Maxa Borna 9,\\ 50-204 Wroclaw, Poland

 \vskip 6mm
 {\small e-mail:\ {gomis-at-ecm.ub.es\\kamimura-at-ph.sci.toho-u.ac.jp
\\lukier-at-ift.uni.wroc.pl}}\\

\medskip

\end{center}

\vskip 20mm
\begin{abstract}
We study  all possible deformations of the Maxwell algebra. In
$D=d+1\neq 3$ dimensions there is only one-parameter deformation.
The deformed algebra is isomorphic  to $ so(d+1,1)\oplus so(d,1)$ or to
$so(d,2)\oplus so(d,1)$ depending on the signs of the deformation
parameter. We construct in the $dS (AdS)$ space a model of massive particle
interacting with Abelian vector field via
non-local Lorentz force. 
In \Deq21  the deformations depend on two parameters $b$ and $k$. We
construct a phase diagram, with two parts of the $(b,k)$ plane with
$ so(3,1)\oplus so(2,1)$ and $ so(2,2)\oplus so(2,1)$ algebras separated by
a  critical curve along which the algebra is isomorphic to
${\Iso}(2,1)\oplus so(2,1)$.
We introduce in \Deq21  the Volkov-Akulov type model for a
Abelian Goldstone-Nambu vector field described by a non-linear
action containing as its bilinear term the free  Chern-Simons
Lagrangean.

\end{abstract}

\setcounter{page}{1}
\parskip=7pt

\newpage
\section{Introduction}

It is known since 1970, see \cite{Bacry:1970ye}, that the presence
of a constant classical EM field background in Minkowski space-time
leads to the modification of Poincare symmetries. One obtains the
enlargement of Poincare algebra, called Maxwell algebra
\cite{Schrader:1972zd},\cite{Beckers:1983gp} which is obtained by
the replacement of the commutative momentum generators $P_a,(a=0,1,...,d)$ by
\be \left[P_{a},~P_{b}\right]= i~e Z_{{ab}}, \quad\quad
Z_{ba}=-Z_{{ab}},\label{NHm1} \ee
where $e$ is the electromagnetic coupling constant.

It is known that the Poincare  algebra does not permit any
central extensions in $D=d+1\, (d>1$) dimensions,
see for example \cite{galindo}.
The new generators $Z_{{ab}}$ describe so called tensorial central
charges\footnote{In euclidean spaces $R^{2n}$ and $R^{4n}$
 with automorphism groups $U(n)$ and $Sp(n)\times SU(2)$ (K\"{a}hler and
hyper-K\"{a}hler geometries), Galperin et.al.
\cite{Galperin:1987wc,Galperin:1987wb} have obtained scalar as well
as tensorial central extensions (a triplet in hyper-K\"{a}hler). In
the literature the tensorial central charges were introduced mostly
in the Poincare superalgebras \cite{D'Auria:1982nx}
\cite{vanHolten:1982mx} \cite{de Azcarraga:1989gm} and also in
p-brane non-relativistic  Galilei and Newton-Hooke algebras
\cite{Brugues:2004an} \cite{Gomis:2005pg} \cite{Brugues:2006yd}.}
 and satisfy the relations
\bea
\left[M_{ab},~Z_{cd}\right]&=&-i~(\h_{bc}~Z_{ad}-\h_{bd}~Z_{ac}+\h_{ad}~
Z_{bc}-\h_{ac}~Z_{bd}),\nn\\
\left[P_a,~Z_{bc}\right]&=&0, \quad\quad
\left[Z_{ab},~Z_{cd}\right]=0 .\label{Poincare0} \eea  A dynamical
realization of Maxwell algebra can be obtained by considering the
relativistic particle coupled in minimal way to the electromagnetic
{ potential} $ A_b=\frac12{f^0_{ab}}x^a$ defining the constant field
strength $F_{ab}={f^0_{ab}}$. The respective first order lagrangian
is the following \be L={\pi}_a\dot x^a-\frac{\lambda}{2}
({\pi}^2+m^2)+\frac{e}2 f_{ab}^0x^a\dot x^b.\label{lagrangian1}\ee
The coordinate ${\pi}_a$ can be expressed in terms of the canonical
momenta $p_a$ conjugated to $ x^a$ as \be
{\pi}_a=p_a+\frac{e}{2}{f^0_{ab}}x^b.\ee From \bref{lagrangian1} we
get the second order lagrangian \be\label{bacry2} L=-m\sqrt{-\dot
x^2}+\frac e2 {f^0_{ab}}x^a\dot x^b.\ee Note that this action is not
invariant under the whole Maxwell algebra since part of the  Lorentz
rotations is  broken by the choice of constant electromagnetic field
${f^0_{ab}}$. In order to recover the Maxwell symmetry one has to
promote ${f^0_{ab}}$ to be the dynamical degrees of freedom
 and  consider an extension of space-time by supplementing the new
coordinates $\theta^{ab} (=-\theta^{ba})$ which are canonically
conjugated to $Z_{ab}$. In order to 
introduce the dynamics invariant under the Maxwell group symmetries
we have applied in \cite{Bonanos:2008kr} \cite{Bonanos:2008ez}  the
method of non-linear realizations employing the Maurer Cartan (MC)
one-forms (see e.g. \cite{Coleman},\cite{Gomis:2006xw}).

The aim of this paper is to study all possible deformations of the
Maxwell algebra \bref{NHm1}, \bref{Poincare0}, and investigate
the dynamics realizing the deformed Maxwell symmetries.

 In  \Dneq21 there exists only one-parameter deformation which
leads for positive (negative) value of the deformation parameter $k$
to an algebra that is isomorphic to the direct sum of the AdS
algebra $so(d,2)$ (dS algebra $so(d+1,1)$) and the Lorentz algebra
$so(d,1)$. We stress here that this deformation for $k>0$ has been
firstly obtained by Soroka and Soroka who called the Maxwell algebra
as the tensor extension of Poincare algebra\cite
{Soroka:2004fj},\cite{Soroka:2006aj}.

 In \Deq21  one gets a two-parameter family of deformations, with
second deformation parameter $b$. The parameter space $(b,k)$ is
divided in two regions separated by the critical curve
\be A(b,k)={(\frac
k3)}^3-{(\frac b2)}^2=0 \label{degencurv}\ee on which the deformed
algebra is non-semisimple. It appears that for $A>0 ~ (A<0)$
the deformed algebra is isomorphic to $so(2,2)\oplus so(2,1)\;
(so(3,1)\oplus so(2,1))$. On the curve \bref{degencurv} the deformed
algebra is  the direct sum of \Deq21  Poincare algebra and
\Deq21  Lorentz algebra, ${\Iso}(2,1)\oplus so(2,1)$.

In order to study the particle dynamics in the deformed cases we
consider the MC one-forms on the suitable coset of deformed Maxwell group.
Firstly we obtain, for arbitrary $D$ and $k\neq0, b=0$,
the particle model in curved and enlarged space-time $y^A=(x^a, \theta^{ab})$.
We choose the coset which leads to
the metric depending only on the space-time coordinates $x^a$.  We
derive in such a case the particle model in AdS ( for $k>0$) or dS (
for $k<0$) curved space-time with the coupling to Abelian vector
field which generalizes,
 in the theory with deformed Maxwell symmetry,
 the Lorentz force term describing the particle interaction with constant
 electromagnetic field. The Lorentz force in the case studied here
becomes non-local.

In \Deq21  and $k=0, b\neq 0$, we will consider a nonlinear field
theory realization of the deformed Maxwell algebra in six-dimensional
enlarged space $(x^a,\theta^{a}=\frac 12 \epsilon^{abc}
\theta_{bc};\,a,b=0,1,2)$ by assuming that the surface
$\theta^{a}=\theta^{a}(x)$ describes \Deq21  dimensional Goldstone
vector fields\footnote{Such a method was used firstly by Volkov and
Akulov to derive the Goldstino field action \cite{Volkov:1973ix}.}.
 If we postulate the action of Volkov-Akulov type
\cite{Volkov:1973ix} \cite{Zumino:1977av} we shall obtain the field
theory in \Deq21 space-time with a lagrangian  containing  a free
Abelian Chern-Simons term\cite{Schwarz:1978cn},
\cite{Deser:1982vy},\cite{Wilczek:1982rv}.

 The organization of the paper is as follows. In section 2
we review some properties of the Maxwell group and consider the corresponding
particle model. In  section 3 we will present all possible deformations of
Maxwell algebra. In section 4 we construct the deformed particle model for
arbitrary $D$  with $k\neq0, b=0$. 
In section 5 we consider \Deq21  case with $k=0, b\neq 0$ and promote the
group parameters $\theta^a$
to Goldstone fields  $\theta^a(x)$. These Goldstone-Nambu fields will be
described by Volkov-Akulov type action.
In the final section we present a short summary and further outlook.
Some technical details are added in two appendices.

\section{Particle Model from the Maxwell algebra }

In this section we construct a particle model invariant under the complete 
Maxwell algebra.
Such a model can be derived geometrically \cite{Bonanos:2008ez} by
the techniques of non-linear realizations, see e.g. \cite{Coleman},
and by the introduction of new dynamical
coordinates $f_{ab}$ that transform covariantly under the Maxwell group.

Let us consider the coset \cite{Bonanos:2008kr}
\cite{Bonanos:2008ez} \be\label{parametrization2}
g=e^{iP_ax^a}e^{\frac{i}{2}Z_{ab}\theta^{ab}}. \label{cosetg0}\ee
The corresponding Maurer-Cartan (MC) one-forms are \be
\label{omega}\Omega=-i g^{-1}dg= P_a~e^a+\frac12
Z_{ab}~\omega^{ab}+\frac12 M_{ab}~l^{ab}, \label{MCform}\ee
where \be e^a=dx^a, \hspace{15 pt}
\omega^{{ab}}=d\theta^{ab}+\frac12(x^a\;dx^b-x^b\;dx^a), \hspace{15
pt} l^{{ab}}=0. \label{explicit1} \ee

Differential realization of the Maxwell algebra generators is described by
the left invariant vector fields in the extended space-time
$y^A=(x^a, \theta^{ab})$\footnote{We could also have a
realization in terms of right invariant vector fields that generate
the infinitesimal transformations \bref{transformations}.}, which
are dual to the one-forms \bref{explicit1} \cite{Soroka:2004fj},\cite{
Bonanos:2008ez}.

A first order form of the lagrangian for the particle invariant under
the full Maxwell algebra with the coordinates $f_{ab}$ describing new
dynamical coupling can be
written as \cite{Bonanos:2008kr}\footnote{ {In the following we put
the electric charge $e$ equal to $1$ for simplicity.}}
 \be \tilde L={\pi}_a {e^a}_A\dot y^A+
\frac{1}2 f_{ab}\,{\omega^{ab}}_A\dot y^A-\frac{\lambda}{2}
({\pi}^2+m^2),\label{lagrangian2}\ee where \be\label{vielbein}
e^a={e^a}_A~ dy^A,\quad\quad \omega^{ab}={\omega^{ab}}_A~ dy^A,\ee
more explicitly,
\bea {e^a}_b&=&{\delta^a}_b,~\quad{e^a}_{bc}=0,\nn\\
{\omega^{ab}}_c&=& \frac12(x^a {\delta^b}_c-x^b{\delta^a}_c),\quad
{\omega^{ab}}_{cd}= \frac12({\delta^a}_c{\delta^b}_d-
{\delta^a}_d{\delta^b}_c). \eea From the \bref{lagrangian2} we
obtain the second order lagrangian \be\label{emlagrangian}
\CL=-m\sqrt{-\dot
x^2}+\frac12f_{ab}\left(\dot\T^{ab}+\frac12(x^a\dot x^b- x^b\dot
x^a)\right)= -m\sqrt{-\dot x^2}+\hat {A}^*. \label{Lag} \ee {The
Euler-Lagrange} equations of motion are \bea\label{eqmotion1theta}
\dot f_{ab}&=&0,
\\\label{eqmotion1f} \dot
\T^{ab}&=&-\frac12(x^a\dot x^b-x^b\dot x^a),
\\ \label{eqmotion1x} m\ddot x_a&=&f_{ab}\dot x^b, \label{eqmotion1xx}\eea
where we took a proper time gauge in \bref{eqmotion1xx}.
Integration of \bref{eqmotion1theta} gives
$f_{ab}=f^0_{ab}$ and such a solution breaks the Lorentz symmetry
spontaneously into a subalgebra of Maxwell algebra. Substituting
this solution in the equation of motion \bref{eqmotion1xx} we
provide the motion of a particle in the constant electromagnetic field
\cite{Bacry:1970ye},\cite{Schrader:1972zd} described by the lagrangian
\bref{bacry2}.
>From \bref{eqmotion1f} one can conjecture that the
new coordinates $\theta^{ab}$ are related with the angular momenta.

 Notice that the interaction part of the lagrangian \bref{Lag}
defines an analogue of the EM potential $\hat A$ as one-form
 in the extended bosonic space $(x,\theta, f)$ \be \hat
A=\frac{1}2\,{f}_{ab}\,\omega^{ab}\label{hatAMax}\ee
The closed two form \be \hat F=d\8A=\frac{1}2\,{f}_{ab}\,e^a\wedge
e^b +\frac{1}2\,{df}_{ab}\wedge\omega^{ab}\,\ee  is such that the second
term vanishes on shell \bref{eqmotion1theta}. We see that on-shell the
field strength has the constant components ${f}_{ab}$.

\subsection{Phase space realization and Casimir operators}

The infinitesimal symmetries of the lagrangian \bref{emlagrangian}
are\footnote{
Our convention of anti-symmetrization is
$A_{[a}B_{b]}=A_aB_b-A_bB_a.$ }
\bea\label{transformations} P_a&:& \D x^a=\ep^a,\quad \D
\T^{ab}=-\frac12(\ep^a x^b-\ep^b x^a),
\nn\\
M_{ab}&:& \D x^a={\lam^a}_b\,x^b,\quad \D
\T^{ab}={\lam^{[a}}_c\,\T^{cb]},\quad \D
f^{ab}={\lam^{[a}}_c\,f^{cb]},\qquad \lam^{ab}+\lam^{ba}=0,
\nn\\
Z_{ab}&:& \D\T^{ab}= \ep^{ab},\qquad \qquad
\ep^{ab}+\ep^{ba}=0. \label{Maxtrans}\eea The corresponding
Noether canonical generators are \bea
\CP_a&=&-\left(p_a-\frac12\,p_{ab} x^b\right),
\nn\\
\CM_{ab}&=&-\left(p_{[a} x_{b]}+ p_{[ac}{\T_{b]}}^c+
p^f_{[ac}{f_{b]}}^c \right),\nn\\
\CZ_{ab}&=&-\,p_{ab}.
 \label{cgenM}\eea They realize the
Maxwell algebra \bref{NHm1} and \bref{Poincare0}, where
$p_a,p_{ab},p_f^{ab}$ are the canonically conjugated momenta of the
coordinates  $x^a, \theta^{ab}, f_{ab}$.

>From the lagrangian \bref{emlagrangian} we obtain the constraints \bea
\phi&=& 
\frac{1}{2}\left(\pi_a^2+m^2\right)=0,\qquad \qquad
\pi_a\equiv p_a+\frac12\,f_{ab}\,x^b,
\nn\\
\phi_{ab}&=&p_{ab}-f_{ab}=0,\qquad
\nn\\
\phi_f^{ab}&=&p_f^{ab}=0.
\label{constraintMax}\eea The last two are the second class constraints
and are solved by the choice $(f_{ab},p_f^{ab})=(p_{ab},0)$.

\vs
The Hamiltonian becomes \bea \CH&=&\lam\, \phi=\frac{\lam
}{2}\left(\pi_a^2+m^2 \right) \eea
and the
Hamilton equations  are, using $p_{ab}=f_{ab}$,
 \bea \dot x^a&=&\lam\,\pi^a, \qquad
\dot p_a=
\frac{\lam}2\,f_{ac}\,\pi^c, 
\nn\\
\dot \T^{ab}&=&\frac{\lam}{2}\, \pi^{[a}\,x^{b]}, \qquad \dot
f_{ab}=0. \eea It follows \be \dot\pi_a=\lam\,{ f}_{ac}\,\pi^c\,,
\label{dotpi}\ee then the constraints \bref{constraintMax}  and the global 
generators \bref{cgenM} are
conserved.

There are four Casimirs in the Maxwell algebra in four dimensions,
\cite{Schrader:1972zd},\cite{Soroka:2004fj}
\bea C_1&=&\CP_a^2-\CM_{ab}\CZ^{ab},\qquad C_2=\frac12\,\CZ_{ab}^2,
\nn\\
C_3&=&(\CZ\t \CZ),\qquad 
C_4=(\CP^{b}\t \CZ_{ba})^2+ \frac14(\CZ\t\CZ)\,(\CM_{ab}\,\t \CZ^{ab}),
\label{4Casimir}\eea
where  $\7\CZ^{ab}=\frac12\ep^{abcd}\CZ_{cd}$.  The values of the
Casimirs are, using the standard $D=4$ notation (${\bf B}=f^{ij}, {\bf
E}=f^{0i}$), \bea
C_1&=& \pi_a^2=-m^2, \qquad 
C_2=\frac12\,f_{ab}^2={\bf B^2-E^2},
\nn\\
C_3&=&\frac12\,\ep^{abcd}f_{ab}f_{cd}=4\,{\bf B\cdot E},
\quad
C_4=\frac12\,m^2\,f^2+ (\pi_b\,f^{ba})^2 
=m^2({\bf B^2+E^2}). \label{valCasimir}\eea
where in the second term of  $C_4$ we took
a frame  in which $\pi_a=(m,0,0,0)$ and imposed the mass shell
constraint. In more general case of  time-like $\pi_a$, ${\bf B, E}$
are defined relatively to the direction of $\pi_a$,
so that expressions for $C$'s remain the ones given by the formulae
\bref{valCasimir}.

\vs

\subsection{First-quantized theory}

Let us observe from  \bref{constraintMax} that the equation
$\phi=C_1+m^2=0$ represents unique first class constraint in the
model. If we introduce first-quantized theory, in the Schr\"odinger
representation, we obtain the following generalized KG equation,
\be\left[\left(\frac1{i}\frac{\pa}{\pa x^a}+\frac1{2i}x^b
\frac{\pa}{\pa \T^{ab}}\right)^2+m^2\right]\Psi(x^a,\T^{ab})=0. \ee
In general case the remaining three Casimirs $C_2,C_3,C_4$ are not
restricted, however  in order to get the irreducible representation
it is necessary to impose their definite values by three
differential equations
\be
C_j\,\Psi(x^a,\T^{ab})=\lam_j\,\Psi(x^a,\T^{ab}),\qquad (j=2,3,4),
\ee
where
\bea
C_2&=&-\frac12\,\frac{\pa}{\pa \T^{ab}}\frac{\pa}{\pa \T_{ab}},
\qquad
C_3=-\frac12\,\ep^{abcd}\frac{\pa}{\pa \T^{ab}}\frac{\pa}{\pa \T^{cd}},
\nn\\
C_4&=&-\frac{m^2}2\, \frac{\pa}{\pa \T^{ab}}\frac{\pa}{\pa \T_{ab}}
+\left((\frac{\pa}{\pa x^b}+\frac1{2}x^c
\frac{\pa}{\pa \T^{bc}})\frac{\pa}{\pa \T_{ba}}\right)^2.
\label{valCasimirS}
\eea
The constraints $C_j\,=\lam_j\,$ can be incorporated into our particle model
by introducing suitable lagrangian multipliers in \bref{lagrangian2}.

\section{Deformations of Maxwell algebra}
\subsection{General considerations}

In this section we would like to find all possible deformations of
the Maxwell algebra. The problem of finding the continuous
deformations of a Lie algebra can be described in cohomological
terms \cite{levynahas}. We first consider the Lie algebra-valued
Maurer-Cartan (MC) one-form
\be \Omega={-i }\,g^{-1}dg= \lambda^a G_a,\qquad
[G_a,G_b]=i\,C_{ab}^c\,G_c,
\ee where
$G_a$'s are generators of a Lie algebra with the structure constants
$C_{ab}^c$  and $\lambda^a$ is the basis of
left-invariant one-forms. The MC equation, $d\W+i\W\wedge\W=0$, becomes
\be d\lambda^a= -\frac12
C_b{}^a{}_c\, \lambda^b\wedge  \lambda^c,\ee
and describes the Lie algebra in terms of dual forms.

We define the matrix-valued one-form $C^a{}_b= \lambda^c C_c{}^a{}_b$
and  following the notation of \cite{Gibbons:2007iu} we consider the
covariant exterior differential  $D\equiv d+C\wedge $  with $D\wedge D=0$.
The infinitesimal deformations are characterized
by the non-trivial vector-valued  two-forms $A^{(2)}$ verifying
\be DA^{(2)} =0\,,\qquad A^{(2)} \ne -D\Phi^{(1)}.  \label{nontriv} \ee
Therefore the non-trivial infinitesimal deformations are  in one to one
correspondence with the second cohomology group $H^2( g;
 g)$.   If a non-trivial linear (infinitesimal)  deformation $A^{(2)}$
is found, the next step is to investigate the Jacobi identities of
higher order in the deformations parameters. The quadratic and
higher deformations are controlled by the cohomology $H^3( g; g)$.
In the case when $H^3( g; g)$ vanishes, it is always possible to
choose a representative in  the class of infinitesimal deformations
such  that it verifies the Jacobi identity in all orders.

Let us apply the above  ideas to the Maxwell algebra \bref{NHm1}. The MC form
for the Maxwell algebra is
\be \Omega=P_a L_P^a+\frac 12 Z_{ab}L_Z^{ab}+\frac 12 M_{ab} L_M^{ab}\ee
The MC equations in this case are given by\footnote{As usual we will
often omit "$\wedge$" for exterior product of forms.}
  \bea
dL_M^{ab}+L_M^{ac}{L_{Mc}}^b&=&0,
\nn\\
dL_P^{a}+L_M^{ac}{L_{Mc}}&=&0,
\nn\\
dL_Z^{ab}+L_Z^{ac}{L_{Mc}}^b+L_M^{ac}{L_{Zc}}^b-L_P^aL_P^b&=&0.
\label{defM}\eea

Expanding the vector-valued two-form $A^{(2)}$ on the basis of one-forms
$L$'s and solving the linear equations resulting from
(\ref{nontriv}) we find a one-parameter family of non-trivial
solutions for $A^{(2)}$, with the exception that there is a
two-parameter family in "exotic" case \Deq21.\footnote{ Some of the
calculations with forms are being done using the Mathematica code
for differential forms EDC \cite{bonanos}.} Infinitesimal
deformations found in this way are not unique but have an ambiguity
described by  $D\Phi^{(1)}$. Since  $H^3( g; g)$ vanishes\footnote{ We
acknowledge Sotirios Bonanos for discussions on this point} finite
deformations are found by adjusting the trivial one-form in a way
providing 
the Jacobi identities for finite values of the deformation parameters.
We find that for any dimension $D$
there is a one-parameter family of exact Lie algebras,
but for \Deq21 there exists a two-parameter family. 
The MC equations get additional terms representing deformations as follows
\bea dL_M^{ab}+L_M^{ac}{L_{Mc}}^b&=&b\,\ep^{abc}L_{Zcd}L_P^d,
\nn\\
dL_P^{a}+L_M^{ac}{L_{Pc}}&=&k\,L_Z^{ac}{L_{Pc}}+b\,\frac14L_Z^{ab}\ep_{bcd}L_Z^{cd},
\nn\\
dL_Z^{ab}+L_Z^{ac}{L_{Mc}}^b+L_M^{ac}{L_{Zc}}^b-L_P^aL_P^b&=&k\,L_Z^{ac}{L_{Zc}}^b,
\qquad
 (\ep^{012}=-\ep_{012}=1).
\label{defQ2}\eea
Here $k$ and $b$ are arbitrary real constant
parameters; we stress that deformation terms proportional to $b$  
are present only  in \Deq21.
The length dimensions of $k$ and $b$ are respectively $[L^{-2}]$ and
$[L^{-3}]$.
In next two subsections we will study these continuous
deformations  using the Lie algebra generators.

\subsection{Arbitrary dimensions}

The general deformed Maxwell algebra found in the previous
subsection can be written in terms of the commutators of generators.
In general dimensions there exists only the following $k$-deformed algebra, 
with $b=0$
\bea
\left[P_a,~P_b \right]&=& i~{Z}_{ab},\qquad
\left[M_{ab},~M_{cd}\right]=-i~\h_{b[c}~M_{ad]}+i~\h_{a[c}~M_{bd]},\nn\\
\left[P_a,~M_{bc}\right]&=&-i~\h_{a[b}~P_{c]}, \qquad
\left[Z_{ab},~M_{cd}\right]=-i\left(~\h_{b[c}~Z_{ad]}-~
\h_{a[c}~Z_{bd]}\right),
\nn\\ 
\left[P_a,~Z_{bc}\right]&=&+ik\,\h_{a[b}~P_{c]},
\nn\\
\left[Z_{ab},~Z_{cd}\right]&=&+ik\left(~\h_{b[c}~Z_{ad]}-i~\h_{a[c}~Z_{bd]}
\right). 
\label{defMaxalg}
\eea
For $k\neq0$ case we introduce dimensionless rescaled generators as
\be
\CP_{a}=\frac{P_{a}}{\sqrt{|k|}},\qquad
\CM_{ab}=-\frac{Z_{ab}}{k},\qquad \CJ_{ab}=M_{ab}+\frac{Z_{ab}}{k},
\label{kdefgene}\ee then the $k$-deformation of Maxwell algebra
becomes
 \bea
\left[\CP_a,~\CP_b\right]&=&-i\,\frac{k}{|k|}\,~{\CM}_{ab}, \nn\\
\left[\CP_a,~\CM_{bc}\right]&=&-i~\h_{a[b}~\CP_{c]},\qquad%
\left[\CM_{ab},~\CM_{cd}\right]=-i~\h_{b[c}~\CM_{ad]}+i~\h_{a[c}~\CM_{bd]},
\nn\\ 
\left[\CP_a,~\CJ_{bc}\right]&=&\left[\CM_{ab},~\CJ_{cd}\right]=0,\qquad
\left[\CJ_{ab},~\CJ_{cd}\right]=-i~\h_{b[c}~\CJ_{ad]}+i~\h_{a[c}~\CJ_{bd]}.
\label{kdefalge}  \eea

The algebra of $(\CP_a,~\CM_{cd},~\CJ_{cd},)$ for $k>0$
($k^+$-deformation) is $\so(D-1,2)\oplus \so(D-1,1)$, {\it i.e.}
we obtain the direct sum of $AdS_D$ and $D$-dimensional Lorentz group. For $k<0$
($k^-$-deformation) we get $\so(D,1)\oplus \so(D-1,1)$, i.e., the direct
sum of $dS_D$ and $D$-dimensional Lorentz group. We recall here that the
above algebra for $k>0$ was previously found by Soroka and Soroka
\cite{Soroka:2006aj}. In our further discussion we will also use the
notation $k=\pm\frac1{R^2}$ where $R$ is the radius of AdS
($k>0$) or dS $(k<0)$ space.

\subsection{\Deq21 }

This  case is interesting since there is an exotic $b$-deformation
of the Maxwell algebra in addition to the $k$-deformation. In \Deq21
it is convenient to use the dual vectors for anti-symmetric tensors,
{\it i.e.}
 \bea M^a&=&\frac12\ep^{abc}M_{bc},\qquad
 Z^a=\frac12\ep^{abc}Z_{bc},\qquad etc.
\eea
The algebra \bref{defMaxalg} looks as follows
\bea
\left[P_{a},P_{b}\right]&=&-i\ep_{abc}Z^c,\qquad
\left[M_{a},M_{b}\right]=i\ep_{abc}M^c,
\nn\\
\left[P_{a},M_{b}\right]&=&i\ep_{abc}P^c,\qquad
\left[Z_{a},M_{b}\right]=i\ep_{abc}Z^c,
\nn\\
\left[P_{a},Z_{b}\right]&=& -ik\,\ep_{abc}P^c-ib\,\ep_{abc}M^c,\qquad
\nn\\
\left[Z_{a},Z_{b}\right]&=&-i k\,\ep_{abc}Z^c+i\,b\,\ep_{abc}P^c.
\label{D3dMax1}\eea

For $b=0,\;k\neq 0$ , as was discussed previously,
the algebra is $\so(2,2)\oplus\so(2,1)$ for $k>0$ ($k^+$-deformation) and
$\so(3,1)\oplus\so(2,1)$ for $k<0$ ($k^-$-deformation).
We rewrite the formula \bref{kdefgene} in a matrix form as
\bea
\pmatrix{\CP_a \cr \CM_a \cr \CJ_a}&=&
\pmatrix{&&\cr& U_k&\cr&& }\pmatrix{P_a \cr M_a \cr Z_a},\qquad
U_k=\pmatrix{\frac{1}{\sqrt{|k|}}&0&0\cr0&0&-\frac1{k}\cr0&1&\frac1{k}}.
\label{transkdef}\eea
The algebra (\ref{kdefalge}) becomes
\bea
\left[\CP_{a},\CP_{b}\right]&=&i\,\frac{k}{|k|}\,\ep_{abc}\CM^c,
\quad
\left[\CP_{a},\CM_{b}\right]= i\,\ep_{abc}\CP^c,
\quad
\left[\CM_{a},\CM_{b}\right]=i \,\ep_{abc}\CM^c,\quad
\nn\\ 
\left[\CJ_{a},\CJ_{b}\right]&=&i\ep_{abc}\CJ^c,
\quad
 \left[\CP_{a},\CJ_{b}\right]=\left[\CM_{a},\CJ_{b}\right]=0.
\label{D3dMax3}\eea

For $k=0,\;b\neq 0$ ($b$-deformation)
we can introduce
\bea
\pmatrix{\CP_a \cr \CM_a \cr \CJ_a}&=&
\pmatrix{&&\cr& U_b&\cr&& }\pmatrix{P_a \cr M_a \cr Z_a},\qquad
U_b=
\pmatrix{\frac1{\sqrt3\,{b^{1/3}}}&0&\frac1{\sqrt3\,{b^{2/3}}}\cr
-\frac1{3\,{b^{1/3}}}&\frac23&\frac1{3\,{b^{2/3}}}\cr
\frac1{3\,{b^{1/3}}}&\frac13&-\frac1{3\,{b^{2/3}}}}
\label{transbdef}\eea
and show that
\bea
\left[\CP_{a}, \CP_{b}\right]&=&-i\,\ep_{abc}\CM^{c}, \quad
\left[\CP_{a}, \CM_{b}\right]=i\ep_{abc}\CP^{c},\quad
\left[\CM_{a}, \CM_{b}\right]=i\ep_{abc}\CM^{c},
\nn\\
\left[\CJ_{a}, \CJ_{b}\right]&=&i\ep_{abc}\CJ^{c},\quad
\left[\CP_{a}, \CJ_{b}\right]=
\left[\CM_{a}, \CJ_{b}\right]=0.\qquad
\label{J1J2alg}\eea
Then $(\CP_a,\CM_a)$ are the $\so(3,1)$ generators and $\CJ_a$ describes
$\so(2,1)$. This algebra is isomorphic to the one with
$b=0, k<0$ ($k^-$-deformation) \bref{D3dMax3}.

To examine more general case with any values of the deformation
parameters $(b,k)$ we consider the Killing form of the algebra
(\ref{D3dMax1}), \bea g_{ij}=C_{ik}^\l C_{\l j}^k&=&6
\pmatrix{1&0&-\frac{2 {k}}{3}\cr
     0&\frac{2 {k}}{3}&- {b}\cr
    - \frac{2 {k}}{3}&- {b}& \frac{2{ {k}}^2}{3} } \otimes
\pmatrix{1&0&0\cr0&-1&0\cr0&0&-1}.
 \label{KFmat3}  \eea
where $C_{ij}^k $ is the structure constant in the base of nine
generators $G_i=(P_a,M_a,Z_a)$.
Its determinant is \be \det g_{ij}=6^94^3\,A(b,k)^3,\qquad
A(b,k)\equiv (\frac{k}{3})^3-\,(\frac{b}2)^2.
\label{degenerate}\ee
In the case $\det g=0$  the Killing form is degenerate,
otherwise the algebra is semisimple.

In figure 1 we illustrate the parameter plane $(b,k)$ which is
divided into four regions in the table 1;
\begin{figure}[hc]
  \begin{center}
\fbox{\includegraphics{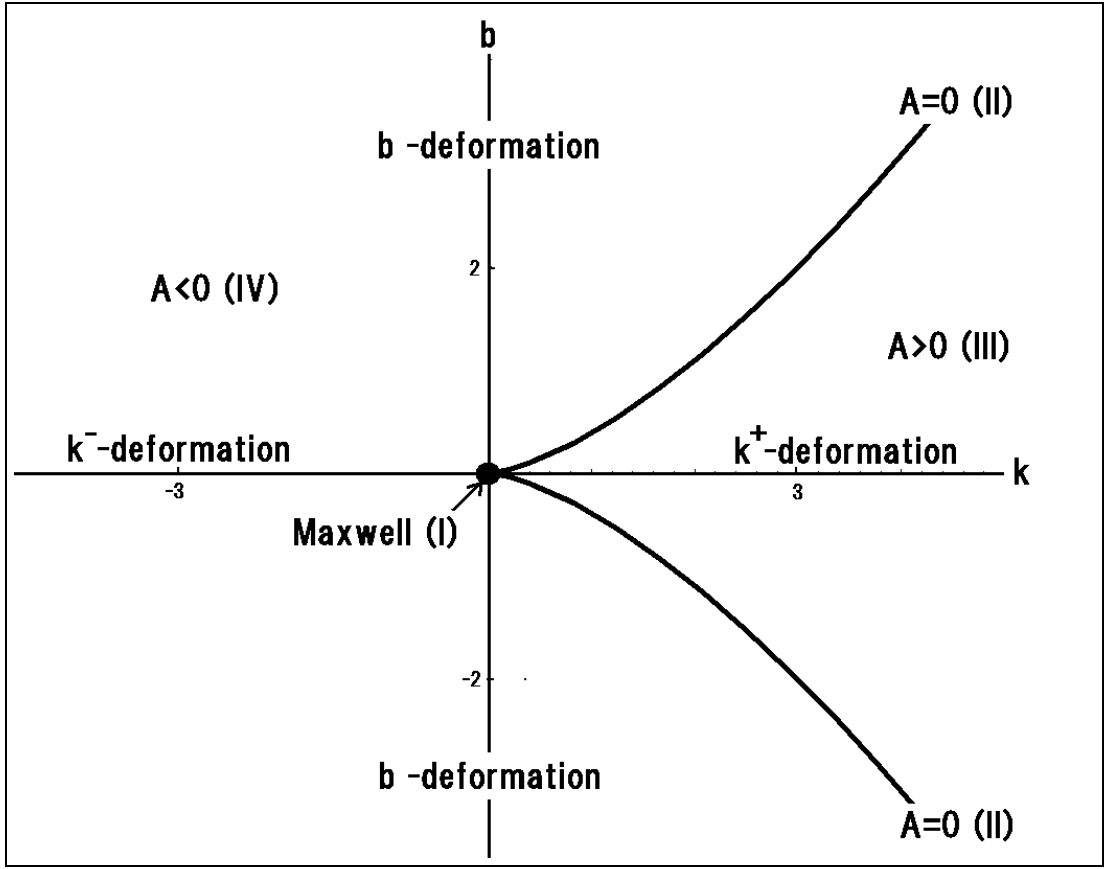}}
 \end{center}
\vskip -6mm
  \caption{{\it The phase diagram for deformed \Deq21 Maxwell algebra}
 }
\end{figure}
  \begin{center}
\begin{tabular}{|c|c|c|c|c|}
  \hline
 &  &  & &  \\
  I & $\det g=0$ & $ b=0,\; k=0,$& {Maxwell} & Maxwell algebra \\
 &  & & &  \\
 II & $\det g=0$ & $A(b,k)=0$& Poincar\'e&
${\Iso}(2,1)\oplus\so(2,1)$ \\
   &  & & &  \\
III & $\det g>0$ & $A(b,k)>0$ & AdS & $\so(2,2)\oplus\so(2,1)$ \\
   &  & & &  \\
IV & $\det g<0$ &   $A(b,k)<0$& dS &$\so(3,1)\oplus\so(2,1)$\\
 &  &  &  & \\
  \hline
\end{tabular}

Table 1:
{\it The phase sectors for deformed \Deq21 Maxwell algebra}
\end{center}
The origin, $b=k=0$ on the figure (see I) is the case of original Maxwell
algebra. When  $(b,k)$ belongs to one of the two branches ($b>0, k>0$ and
$b<0, k>0$) of the degenerate curve (see II)
we find that the algebra is a direct sum of \Deq21 Poincare
$({\Iso}(2,1))$ and $so(2,1)$. The generators are \bea
\pmatrix{\CP_a \cr \CM_a \cr \CJ_a}&=&
\pmatrix{(\frac2{b})^{1/3}&2&(\frac2{b})^{2/3}\cr
-\frac29\,(\frac2{b})^{1/3}&\frac89&\frac19\,(\frac2{b})^{2/3}\cr
\frac29\,(\frac2{b})^{1/3}&\frac19&-\frac19\,(\frac2{b})^{2/3}}
\pmatrix{P_a \cr M_a \cr Z_a}
\label{dengene}\eea
and satisfy
\bea
\left[\CP_a,\CP_b\right]&=&0,\quad
\left[\CP_a,\CM_b\right]=i\ep_{abc}\CP^c,\quad
\left[\CM_a,\CM_b\right]=i\ep_{abc}\CM^c,\quad
\nn\\
\left[\CJ_a,\CJ_b\right]&=&i\ep_{abc}\CJ^c,\quad
\left[\CP_a,\CJ_b\right]=\left[\CM_a,\CJ_b\right]=0.
\label{dengenealge}\eea

\vs

The AdS region (see III) includes $k^+$-deformation for which the algebra is
$\so(2,2)\oplus\so(2,1)$. One can show that the deformed algebra
for any internal point  $(b,k)$ in sector (III) is isomorphic to
$\so(2,2)\oplus\so(2,1)$.
The generators are constructed as linear combinations of $(P,M,Z)$,
\bea
\pmatrix{\CP_a \cr \CM_a \cr \CJ_a}&=&
\pmatrix{&&\cr& U^+(b,k)&\cr&& }
\pmatrix{P_a \cr M_a \cr Z_a},\qquad U^+(b,k)\in GL(3,R)
\label{transUp}\eea
and $(\CP_a,\CM_a,\CJ_a)$ verify the same AdS algebra as \bref{D3dMax3}
with $k>0.$
The explicit form of the matrix $U^+(b,k)$ is discussed in the Appendix.
It is ill-defined as $(b,k)$ approaches the boundary (II)
because of the appearance of singular coefficients.

The dS region (see IV) includes $k^-$- 
and $b$-deformations for which the algebra is isomorphic
to $\so(3,1)\oplus\so(2,1)$. It is true for general deformations
corresponding to any internal point $(b,k)$ in sector (IV), 
\bea
\pmatrix{\CP_a \cr \CM_a \cr \CJ_a}&=&
\pmatrix{&&\cr& U^-(b,k)&\cr&& }
\pmatrix{P_a \cr M_a \cr Z_a},\qquad U^-(b,k)\in GL(3,R),
\label{transUm}\eea
and $(\CP_a,\CM_a,\CJ_a)$ verify the same dS algebra as 
\bref{D3dMax3}
with $k<0,$ or equivalently given by \bref{J1J2alg}.
It is shown in appendix A that when $k=0$ the transformation matrix
$ U^-(b,k)$ is $U_b$ described by \bref{transbdef},
while it becomes $U_k$ of $k^-$-deformation \bref{transkdef} when $b=0,k<0$.
It is singular as $(b,k)$ approaches the boundary (II), similar as in
region (III).
\vs

Finally we observe that if a new length scale $R'$
is introduced by the relation
\be b=\frac1{R^{'3}} \ee the critical curve equation $A=0$ (see
\bref{degencurv}) is described by two
half-lines relating the parameters $R$ and $R'$, \be
R'=\pm\,\left(\frac{3^{\frac12}}{2^{\frac13}}\right)\, R,\quad
(R>0). \ee

\section{Particle models on the $k$-deformed Maxwell \\ algebra }

In this section we will discuss a model realizing in arbitrary dimension $D$
the deformed Maxwell algebras and look for the physical meaning of the
additional coordinates $(f_{ab},\T^{ab})$.
Using the techniques of non-linear realization, (see e.g. \cite{Coleman}),
 we generalize the results described in sect 2 for the standard Maxwell algebra
\cite{Bonanos:2008ez}
to those for the deformed Maxwell algebra with $(k\neq0,b=0)$. 
In such a way we obtain the generalization to AdS(dS)
space-time of the model describing the particle interacting with constant
values of electromagnetic field via the Lorentz force.

\subsection{Standard Parametrization of the Coset}
 We consider a coset ${G}/{H}$ with $G=\{P_a,M_{ab},Z_{ab}\},
H =\{M_{ab}\}$ and parametrize the group element $g$ using
 $(x^a,\theta^{ab})$, the group parameters associated to the
generators $(P_a,Z_{ab})$. Following to \bref{cosetg0}  we define
\be
g=e^{iP_ax^a}e^{\frac{i}{2}Z_{ab}\theta^{ab}}.
\label{cosetg}\ee
The space-time symmetry is $k$-deformed Maxwell algebra \bref{defMaxalg}
and $\{P_a,Z_{ab}\}$ form  subalgebra  generators isomorphic to
those of AdS (dS) for $k>0, (k<0)$.
The MC form for this coset is
\bea \W&=&-ig^{-1}dg=L_P^aP_a+\frac12L_Z^{ab}Z_{ab}+\frac12L_M^{ab}M_{ab},
\label{MC}\eea
where 
\bea
L^a_P&=&{e}^b\,{\Lambda_b}^a,\quad
L_Z^{cd}=-\frac1{k}\,{\Lambda^{-1c}}_a\,\left(\w^{ab} -({\Lambda}\,d
{\Lambda^{-1}})^{ab}\right){\Lambda_b}^d,\quad
L_M^{cd}=0,\label{MCForms}\eea
and  ${\Lambda}$ is a vector Lorentz transformations (Lorentz harmonics) 
in terms of  new tensorial coordinates $\T^{ab}$ as follows
\be
{\Lambda_a}^b={(e^{-k\T})_a}^b={\D_a}^b+{(-k\T)_a}^b+\frac1{2!}{(-k\T)
_a}^c{(-k\T)_c}^b+...\, .\label{Lamab}\ee
Remember the indices $a,b,...$ are rised and 
lowered using the Lorentz metric $h_{ab}=(-:+...+)$.
One-forms $(e^a,\w^{ab})$ are
\bea e^a&=& dx^c {e_c}^a=dx^c\,\left({\D_c}^a+
(\frac{\sin(\sqrt{kr^2})}{\sqrt{kr^2}}-1)({\D_c}^a-\frac{x_cx^a}{x^2})\right),
\nn\\
\w^{ab}&=& dx^c {\w_c}^{ab}= dx^c \,{\frac{{\D_c}^{[a}\,x^{b]}}{x^2}}(\cos({\sqrt{kr^2}})-1),
\eea where $r=\sqrt{-x^ax_a}$ (for dS case $k=-1/R^2<0$,
$\frac{\sin(\sqrt{kr^2})}{\sqrt{kr^2}}$ is replaced by
$\frac{\sinh(\sqrt{-kr^2})}{\sqrt{-kr^2}}$ correspondingly). 
They satisfy the known AdS
MC equations \be de^a+{\w^a}_be^b=0,\qquad
d\w^{ab}+{\w^a}_c\w^{cb}=-k\,e^ae^b \ee and $L$'s in \bref{MCForms} satisfy the MC
equations \bref{defQ2}, with $b=0$. Remembering $L_M^{ab}=0$ they
are
\bea 
dL_P^{a}&=&k\,L_Z^{ac}{L_{Pc}},\qquad 
dL_Z^{ab}=L_P^aL_P^b+k\,L_Z^{ac}{L_{Zc}}^b.
\label{defQ22}\eea

 The particle action generalizing  \bref{Lag} for $k\neq 0$ looks as follows
\bea
\CL\,d\tau=-m\sqrt{-\h_{ab}\,L_P^{a*}L_P^{b*}}+\frac12f_{ab}L_Z^{ab*}=
-m\sqrt{-g_{ab}(x)\,\dot x^a\dot x^b}\,d\tau+\8A^*,
\qquad  \label{Lag50}\eea
where $g_{ab}$ is the metric, now depending only on $x$,
\bea
g_{ab}&=&{e_a}^c{e_b}^d\h_{cd}=
\h_{ab}+\left[(\frac{\sin(\sqrt{kr^2})}{\sqrt{kr^2}})^2-1\right]
(\h_{ab}-\frac{x_a x_b}{x^2}). \label{metric1}\eea
We obtain the metric of AdS (dS)
space with radius $R$, where $k=1/R^2$. The
pullback $\8A^*$ in \bref{Lag50} takes the explicit form
\bea
\8A^*&=&-\frac12f_{ab}L_Z^{ba*}
=-tr(\frac12f\,L_Z^{*})
=+\frac1{2k}\,tr\left[
f\,\Lambda^{-1}\left( \w_\tau-\Lambda\,\pa_\tau
\Lambda^{-1}\right)\Lambda\right]\,d\tau,
\nn\\
&& \w_\tau^{cd}=\frac{\dot x^{[c}x^{d]}}{{x^2}}\,
(\cos({\sqrt{kr^2}})-1). \label{Lag3}\eea
In a limit $k\to 0$ (equivalently $R\to\infty$) we obtain the
undeformed Maxwell case \bref{hatAMax}\cite{Bonanos:2008ez}.

Now we shall describe the equations of motion following from the
lagrangian \bref{Lag50}.
Taking the variation {with respect to } $f_{ab}$ we get (we suppress the 
tnsor indices)
\bea
\w_\tau-\Lambda\,\pa_\tau
\Lambda^{-1}=0.
\label{eomf0}\eea
In the  limit $k\to 0$  the terms linear in $k$ reproduce the equation \bref{eqmotion1f}.
Comparing with \bref{MCForms}
we see that the pullback of $L_Z^{ab}$ to the world line vanishes on shell.
The variation with respect to $\T^{ab}$ is simplified using \bref{eomf0} and
becomes the same equation as in the Maxwell case \bref{eqmotion1theta}
\bea
\dot f_{ab}&=&0.
\label{eomT0}\eea
Finally variation with respect to $x^a$ gives, after using \bref{eomf0},
the generalization of equation of motion \bref{eqmotion1xx} describing
particle moving under the Lorentz force,
\bea
m\,\nabla_\tau \dot x_a=F_{ab}\dot x^b,
\label{eomx0}\eea
where\bea
\,\nabla_\tau \dot x_a&\equiv&
\frac{g_{ab}}{\sqrt{-g_{ef}\dot x^e\dot x^f}}\left({\sqrt{-
g_{ef}\dot x^e\dot x^f}}\,\pa_\tau\frac
{\dot x^b}{\sqrt{-g_{ef}\dot x^e\dot x^f}}
+{\Gamma^b}_{cd}\dot x^c\dot x^d\right),
\nn\\ 
{\Gamma^b}_{cd}&=&\frac12g^{ba}(g_{ac,d}+g_{ad,c}-g_{cd,a}),
\eea and \bea
F_{ab}(x,\theta)&=&(\Lambda\,f\,\Lambda^{-1})_{cd}\,{e_a}^c\,{e_b}^d\,
\nn\\
&=&\7f_{ab}\,
\left(\frac{\sin(\sqrt{kr^2})}{\sqrt{kr^2}}\right)^2-
\frac{\7f_{[ac}x^cx_{b]}}
{x^2}\,\frac{\sin(\sqrt{kr^2})}{\sqrt{kr^2}}\,
\left(\frac{\sin(\sqrt{kr^2})}{\sqrt{kr^2}}-1\right)
\label{Fmn}\eea
provided that  \be \7f_{ab}
=(\Lambda\,f\Lambda^{-1})_{ab}.
\ee
We see that for $k\neq0$ the generalized Lorentz force 
depends on $\T^{cd}$ but  in the limit $k\to 0$ we get $F_{ab}=f_{ab}$
as expected.

The interaction term $\8A^*$ in the lagrangian \bref{Lag50}
defines an analogue of the EM
potential $\8A$ as one-form in the extended bosonic space of
$y^A=(x^a,\T^{ab})$. Due to the MC eq.   \bref{defQ22}  its
field strength is
\be
\8F=d\,\8A=\frac12f_{ab}\,L_P^a\wedge L_P^b+\frac{k}2\,f_{ab}\,L^a_{Zc}\wedge L_Z^{cb}
+\frac12\,(df_{ab})\wedge L_Z^{ab} \label{hatFF}\ee 
The first term depends only on the coordinate differential forms
$dx^a$ and it can be shown from \bref{Fmn} that \be
\frac12f_{ab}L_P^a\wedge L_P^b=\frac12\, F_{ab}\,dx^a\wedge dx^b,
\label{constF}\ee where $F_{ab}$ is given in \bref{Fmn} as
appears in the equation of motion \bref{eomx0}. The second
and third terms of \bref{hatFF} contain  $L_Z$ and $df_{ab}$ whose pullback
vanish as the result of the equations of motion \bref{eomf0} and
\bref{eomT0}.
We can say that field strength occurring in \bref{eomx0} can be 
regarded as the one described by the generalized U(1) gauge
 potential\footnote{
The U(1) gauge transformation is considered in the extended space as
$\8A\to\8A+d\Lambda(x,\T,f)$ under which $\8F$, therefore
$F_{ab}$, remains invariant on-shell.} $\8A$ in the extended
bosonic space $(x^a,\T^{ab},f_{ab})$. We see that on-shell (i.e.
modulo equations of motion) this field strength has only constant
components $f_{ab}$ with respect to the two-form base
$L^a\wedge L^b$. We conclude that on-shell $f_{ab}$ is a constant
(see \bref{eomT0}) and the variables $(x^a(\tau),\T^{ab}(\tau))$
satisfy the set of nonlinear differential equations, \bref{eomf0}
and  \bref{eomx0}. If we express $\T^{ab}(\tau)$ by using
\bref{eomf0} in terms of variables $x^a(\tau)$, and substitute 
into $\Lambda(\T)$ defining $F_{ab}$, we obtain the generalized
Lorentz force, which is  {\it nonlocal} in the variable $x^a(\tau)$.

\vs

\subsection{Second Parametrization of the Coset}

 In this subsection we shall consider
a different choice of the coset parametrization,
 \be
g'=e^{iP_ax^a}e^{\frac{i}{2}Z_{ab}\theta^{ab}}\,h,\qquad
h=e^{\frac{i}{2}kM_{ab}\theta^{ab}}\in H. \label{cosetg2}\ee
The $g'$ and  $g$ differ by an element of $H$ and
are equivalent representatives of the coset element of $G/H$.
In particular in the $k\to 0 $ limit both $g$ and  $g'$  coincide.
Using the basis \bref{kdefgene} we get
 \be
g'=
e^{i\sqrt{|k|}\CP_ax^a}e^{\frac{i}{2}k\CJ_{ab}\theta^{ab}}.
\label{cosetg3}\ee
The  MC one-forms obtained from  \bref{cosetg3} can be expressed in two
bases of $k$-deformed Maxwell algebra, \bref{defMaxalg} and
\bref{kdefalge}, as follows;
\bea
\W'=-ig^{'-1}dg'
&=&L^a_{\CP}\CP_a+\frac12L_\CM^{ab}\CM_{ab}+\frac12L_\CJ^{ab}\CJ_{ab}
\nn\\&=&
L^{'a}_PP_a+\frac12L_Z^{'ab}Z_{ab}+\frac12L_M^{'ab}M_{ab}.
\label{MC3}\eea
The explicit forms of the MC one-forms are
\bea
L^a_\CP&=& {\sqrt{|k|}}\,{e}^a,  \qquad
L_\CM^{ab}=\w^{ab},  \qquad
L_\CJ^{ab}=(\Lambda\,d\,\Lambda^{-1})^{ab}.
\label{MCForms2}\eea
then
\bea L^{'a}_{P}&=&\frac{L_\CP^{a}}{\sqrt{|k|}}=e^a,\quad
 L_M^{'ab}=L_\CJ^{ab}=(\Lambda\,d\,\Lambda^{-1})^{ab},
\nn\\ L_Z^{'ab}&=&\frac1{k}(L_\CJ^{ab}-L_\CM^{ab})=
-\frac1{k}(\w^{ab}-(\Lambda\,d\,\Lambda^{-1})^{ab})).
\label{MCkdefgp} \eea
Note that $L^a_\CP$  and $L_\CM^{ab}$  are given by the vielbein and
the spin connection of the AdS (dS) space  and $L_\CJ^{cd}$ is the spin connection of the "external" Lorentz space.
We can interpret $L_Z^{'ab}$ as the difference of these spin connections.

 The particle action on the coset \bref{cosetg2}  invariant
under the deformed Maxwell algebra can be obtained by replacing $L$ by
$L'$ in the action \bref{Lag50}. We get
  \bea
\CL'\,d\tau=-m\sqrt{-\h_{ab}\,L_P^{'a*}L_P^{'b*}}+\frac12f'_{ab}L_Z^{'ab*}=
-m\sqrt{-g'_{ab}(x)\,\dot x^a\dot x^b}\,d\tau+\8A^{'*},
\qquad \label{Lag5}\eea where   $g'_{ab}$ is same AdS(dS) metric
 \bref{metric1}  obtained  in the previous parametrization, \bea
g'_{ab}(x)&=&{e_a}^c{e_b}^d\h_{cd}
=g_{ab}(x). \label{metric2}\eea
The interaction term  $\8A^{'*}$ is written in terms of an auxiliary dynamical
variable $f'_{ab}$ as
\bea \8A^{'*}&=& -\frac12f'_{ab}L_Z^{'ba*}=
- \frac12\,tr(f'L_Z{'})=
 \frac1{2k}\,tr \left[f'( \w_\tau-\Lambda\,\pa_\tau\Lambda^{-1})\right]
\,d\tau. 
\label{Lag32}\eea
Then lagrangian $\CL$ in \bref{Lag50} and $\CL'$ in \bref{Lag5}
can be identified if
\be
f'=\Lambda\,f\,\Lambda^{-1}.
\label{ffd}\ee
Since this is a point transformation of the coordinates, from
$\{x^a,\T^{ab},f_{ab}\}$ to  $\{x^a,\T^{ab},f'_{ab}\}$,
the Euler-Lagrange equations of these lagrangians are equivalent.

Let us calculate the equations of motion which follow from the
lagrangian \bref{Lag5}. Taking the variation with respect to
$f'_{ab}$ we get
\bea  \w_\tau-\Lambda\pa_\tau{\Lambda^{-1}}=0,\qquad
\label{eomf3}\eea It
coincides with \bref{eomf0} and means that the pullback of $L_Z^{'ab}$ to the world line 
vanishes on shell.  In geometrical terms  the "gravitational" AdS spin
connection coincides  on shell with 
the "external " Lorentz spin connection. Using \bref{eomf3} 
the variation of the lagrangian with respect to $\T^{ab}$ is written  as
\be \pa_\tau f'_{ab}+
{(\Lambda\pa_\tau{\Lambda^{-1}})
_{[a}}^cf'_{cb]}=\pa_\tau f'_{ab}+
{\w_{\tau[a}}^cf'_{cb]}\equiv D_\tau f'_{ab}=0.
\label{feq}\ee
If we use the relation \bref{ffd} it gives the same equation as \bref{eomT0}
obtained in the first
parametrization. Finally  equation of motion for $x $ which  define
the  generalized non-local Lorentz force is \bea
m\,\nabla_\tau \dot x_a 
=F'_{ab}\dot x^b,
\label{eomx2}\eea
where
\bea
F'_{ab}&=&f'_{cd}\,{e_a}^c\,{e_b}^d.
\label{Fdmn}\eea
Again  using the relation \bref{ffd} we obtain
\be f'=\7f=\Lambda \,f\,\Lambda^{-1} \qquad {\rm and }\qquad 
F'_{ab}=F_{ab}
\ee
or equivalently 
\be
\8F'=\frac12\,F'_{ab}dx^a\wedge dx^b
=\frac12f'_{ab}\,L_P^{'a}\wedge L_P^{'b}=\frac12f_{ab}\,L_P^{a}\wedge L_P^{b}=\8F.
\label{hatFF2}\ee
We conclude that on-shell (i.e. modulo equations of
motion) the field strength has constant components $f_{ab}$
with respect to the 2-form base $L_P^a\wedge L_P^b$ but in the new base
 $L_P^{'a}\wedge L_P^{'b}$ the variables $f'_{ab}$ are covariantly constant
(see eq.\bref{feq}).

It will be interesting to have a  physical interpretation of the non-local
Lorentz force in AdS(dS) space (see \bref{eomx0} and \bref{eomx2}).

\section{$b$-deformed Maxwell algebra in \Deq21  and \\
Goldstone-Nambu vector fields}

In  \Deq21   one can introduce second deformation parameter $b$.
In this section after the calculation of the MC one-forms for the
$b$-deformed coset space we shall use the resulting geometry to introduce an
action for \Deq21  Abelian Goldstone-Nambu fields.

\subsection{  MC forms for $b$-deformed Maxwell algebra in  \Deq21}

We consider the coset \bref{cosetg} in \Deq21 in order to define the extended space-time
$(x^a,\T_{a}$) for the algebra \bref{D3dMax1}  with $k=0$
\be
g=e^{ix^aP_a}\,e^{i\T^a Z_a}=g_0\,e^{i\T^a Z_a},\qquad g_0=e^{ix^aP_a}.
 \label{cosetg5} \ee
We compute the MC one-form in two steps,
\be
\W=-ig^{-1}dg=e^{-i\T^a Z_a}\,\W_0\,e^{i\T^a Z_a}
-i\,e^{-i\T^a Z_a}\,d\,e^{i\T^a Z_a}.
\label{Omegabb}\ee
Firstly we calculate
\bea \W_0=g_0^{-1}dg_0=L_{0P}^aP_a+L_{0Z}^{a}Z_{a}+L_{0M}^{a}M_{a},\eea
where
\bea
\pmatrix{L_{0P}^a\cr L_{0M}^a\cr L_{0Z}^a}&=&
\left[ \pmatrix{1\cr 0\cr 0}{\D^a}{}_c+
\pmatrix{F_0(Y)-1\cr bx^2\,F_2(Y)\cr-b(x^2)^2\,F_4(Y) }
{O^a}{}_c+\pmatrix{-bx^2\,F_3(Y)\cr-b^2(x^2)^2\,F_5(Y) \cr F_1(Y) }
{\ep^{a}}_{cb}x^b \right]\,dx^c.
\nn\\ \label{L0pmz}\eea
Here ${O_{a}}^b=({\D_{a}}^b-\frac{x_ax^b}{x^2})$ and
\be
F_i(Y)
=\,\sum_{n=0}\frac{Y^{6n}}{(6n+i+1)!},
\qquad  Y=b^{\frac13} (x^2)^{\frac12},\qquad (i=0,1,2,3,4,5).
\label{funcFi}\ee
The explicit forms of functions $F_i(Y)$'s are given in appendix B.

The complete MC one-form $\W$ becomes
\bea \W=L_{P}^aP_a+L_{Z}^{a}Z_{a}+L_{M}^{a}M_{a},\eea
with
\bea
\pmatrix{L_{P}^a\cr L_{M}^a\cr L_{Z}^a}&=&
\left[ \pmatrix{0\cr 0\cr 1}{\D^a}{}_c+
\pmatrix{-b^3(\T^2)^2\,F_4(Y')\cr -b^2\T^2\,F_2(Y')\cr F_0(Y')-1 }
{\7O^a}{}_c+\pmatrix{-b\,F_1(Y')\cr -b^4(\T^2)^2\,F_5(Y')\cr b^2\T^2\,F_3(Y') }
{\ep^{a}}_{cb}\T^b \right]\,d\T^c
\nn\\ &+&\left[(I_3)\,{\D^a}{}_c+\pmatrix{V_O}{\7O^a}{}_c+\pmatrix{V_E}
{\ep^{a}}_{cb}\T^b \right]\,
\pmatrix{L_{0P}^c\cr L_{0M}^c\cr L_{0Z}^c},
\label{totalLb}\eea
where  $\,
{\7O_{a}}{}^b=({\D_{a}}^b-\frac{\T_a\T^b}{\T^2})$ and $L_0^c$'s are given in
\bref{L0pmz}. $\,I_3, V_O,$ and $V_E$ are
$3\times3$ matrices acting on the three vector
$\pmatrix{L_{0P},L_{0M},L_{0Z}}$.
$\,I_3$ is the unit matrix and
\bea
\pmatrix{V_O}&=&\pmatrix{f_0(Y')-1,&b\T^2\,f_2(Y'),&-b^3(\T^2)^2\,f_4(Y')\cr
b^3(\T^2)^2\,f_4(Y')&f_0(Y')-1,&-b^2\T^2\,f_2(Y')\cr
-b\T^2\,f_2(Y'),&-b^2(\T^2)^2,\,f_4(Y')&f_0(Y')-1},
\nn\\
\pmatrix{V_E}&=&\pmatrix{b^2\T^2\,f_3(Y'),&b^3(\T^2)^2\,f_5(Y'),&-b\,f_1(Y')\cr
b\,f_1(Y'),&b^2\T^2\,f_3(Y'),&-b^4(\T^2)^2\,f_5(Y')\cr
-b^3(\T^2)^2\,f_5(Y'),&-\,f_1(Y'),&b^2\T^2\,f_3(Y')}.
\eea
Here the functions $F_i(Y')$'s are given in \bref{funcFi} and
the functions $f_i(Y')$'s are
\be
f_i(Y')\equiv \sum_{n=0}\frac{{Y'}^{6n}}{(6n+i)!},\qquad
Y'=b^{\frac23}(\T^2)^{\frac12},\qquad (i=0,1,2,3,4,5).
\label{funcfi}\ee
The explicit forms of \bref{funcfi} are listed in appendix B.

For small deformation parameter $b$ the MC one-forms are, 
up to $b^2$,
\bea
L_{P}^a&=& dx^a-b\left((\frac{x^2}{4!}\,{\ep^{a}}_{bc}x^c\,+
\frac12\,{\ep^{a}}_{cd}\T^d\,{\ep^{c}}_{be}x^e\,)dx^b+
\frac12\,{\ep^{a}}_{bc}\T^c\,d\T^b\right)
\nn\\&+&b^2\,\left(
\frac{x^2\,\T^2}{2!3!} \, {\7O^a}{}_c\, \, {O^c}{}_b  +
\frac{\T^2}{3!} \,  {\ep^{a}}_{bc}\T^c\,+
\frac{(x^2)^2}{5!} \,  {\ep^{a}}_{cd}\T^d\,  {O^c}{}_b +
\frac{(x^2)^3}{7!}\,  {O^a}{}_b\right)\,dx^b\,+\CO(b^3),
\nn\\
L_{M}^a &=& b\left(\frac{x^2}{3!}\,{O^a}{}_c\,dx^c+  {\ep^{a}}_{cb}\T^b\,dx^c
\right)\nn\\&-&b^2\,\left(
(
\frac{\T^2}{2!2!} \, {\7O^a}{}_c\, {\ep^{c}}_{bd}x^d +
\frac{x^2}{4!} \,  {\ep^{a}}_{cd}\T^d\,{\ep^{c}}_{be}x^e+
\frac{(x^2)^2}{6!} \,  {\ep^{a}}_{bc}x^c\,)dx^b
\,+ \frac{\T^2}{3!} \, {\7O^a}{}_b\,d\T^b\right)+\CO(b^3),
\nn\\
L_{Z}^a &=&d\T^a +\frac12 {\ep^{a}}_{cb}x^b\,dx^c
-\,b \left(\frac{(x^2)^2}{5!}\,{O^a}{}_c+
\frac{\T^2}{2!} \, {\7O^a}{}_c +
\frac{x^2}{3!} \,{\ep^{a}}_{db}\T^b\, {O^d}{}_c \right)\,dx^c\nn\\&+&b^2\,
\left(\{
\frac{x^2\,\T^2 }{2!4!} \, {\7O^a}{}_c\, {\ep^{c}}_{bd}x^d +
(\frac{(x^2)^2}{6!}+\frac{\T^2}{2!3!})
\,{\ep^{a}}_{cd}\T^d\,{\ep^{c}}_{be}x^e+
\frac{(x^2)^3}{8!} \,  {\ep^{a}}_{bc}x^c\,\}dx^b
\,\right.
\nn\\&&+\left. \frac{\T^2}{4!} \,{\ep^{a}}_{bc}\T^c\,d\T^b\right)
\,+\,\CO(b^3).
\label{MCbexp}\eea
Using the formula \bref{MCbexp} one can calculate the metric in the
extended space-time $y^A=(x^a,\T^{ab})$ with the following decomposition
\bea
g_{AB}(y)\dot y^A \dot y^B
&=&g_{ab}(x,\T)\dot x^a \dot x^b+2\,g_{a{\bar b}}(x,\T)\dot x^a \dot
\T^{\bar b}+g_{{\bar a}{\bar b}}(\T)\dot \T^{\bar a} \dot\T^{\bar b},
\eea
where
\bea
g_{ab}(x,\T)&=&\h_{ab}-b\left( (x\T)\h_{ab}-\frac12(x_a\T_b+x_b\T_a)
\right)
\nn\\&+&b^2
\left((\frac{2}{7!}-\frac{1}{(4!)^2})(x^2)^3(\h_{ab}-\frac{x_ax_b}{x^2})
-\frac{x^2}{5!}(x_a\,\ep_{bcd}x^c\,\T^d+x_b\,\ep_{acd}x^c\,\T^d) \right.
\nn\\&+&
\left.(\frac{1}{4}(x\T)^2+\frac{1}{6}\,x^2\,\T^2)\,\h_{ab}+
\frac{1}{12}\,x^2\,\T_a\T_b-\frac{1}{6}\,(x\T)\,(x_a\T_b+x_b\T_a)
-\frac{1}{6}\,\T^2\,(x_a x_b)\right)+\CO(b^3)
\nn\\
g_{a{\bar b}}(x,\T)&=&-\frac{b}{2}\,\ep_{a{\bar b}c}\T^c+ \frac{b^2}{4}
\left(\frac{x^2}{12}\left(\T_a\,x_{\bar b}-(x\T)\h_{{a}{\bar b}}\right)+
\T^2\,\ep_{a{\bar b}c}x^c\,+ \T_{\bar b}
\,\ep_{acd}x^c\T^d\right)+\CO(b^3),
\nn\\
g_{{\bar a}{\bar b}}(\T)&=&-b^2\,
\frac{\T^2}{4}\,\left( \h_{{\bar a}{\bar b}}-\frac{\T_{\bar a}\T_{\bar b}
}{\T^2}\right)
+\CO(b^3).
\label{metricbb}
\eea
It can be checked from the general formula \bref{totalLb} for $L_P^a$
that in all orders of $b$ the metric
$g_{{\bar a}{\bar b}}$ does not depend on the space-time coordinates $x^a$.

\subsection{Nonlinear action for \Deq21  Goldstone-Nambu vector fields}

In order to introduce the \Deq21 Goldstone-Nambu fields $\theta^a(x)$
we replace the coset \bref{cosetg5} describing the
coordinates $(x^a,\theta^a)$ in our generalized \Deq21 space-time by
\be\label{parametrization3}
\7g=e^{iP_ax^a}e^{{i}Z_{a}\theta^{a}(x)}.\ee
Here the independent coordinates are $x^a$, and the fields $\T^a(x)$ describe
three-dimensional submanifold in $(x^a,\T^a)$.
The fields $\theta^{a}(x)$ transform homogeneously under the
so(2,1) rotations generated by $M_{a}$, but inhomogenously under the
generators $Z_{a}$, what implies spontaneous
breaking of $Z_{a}$ symmetries. The Goldstone-Nambu fields describing
spontaneously broken directions in extended space-time were
introduced by nonlinear realization method \cite{Volkov:1973vd}
\cite{Ivanov:1975zq} in supersymmetric theories.
The broken directions were provided by
the odd superspace degrees of freedom describing fermionic
Goldstino fields \cite{Volkov:1973ix}, or by introducing in $D$
dimensional space-time the $p$-brane fields ($D>p+1$)  ( spontaneously
broken directions are transversal to the p-brane, see for example
\cite{Henneaux:1984mh} \cite{Hughes:1986fa} \cite{Gomis:2006xw}). In
this section we shall convert in \Deq21 the additional degrees of freedom
$\theta^{a}$ into Abelian vectorial Goldstone-Nambu fields
$\theta^{a}(x)$ which can be also interpreted as describing a
3-brane in $D=6$ space time $(x^a, \theta^{a})$.

In order to study the dynamics of fields $\theta^{a}(x)$ we should
calculate, using \bref{parametrization3}, the left-invariant MC
one-forms
\be \label{omegava}\tilde\Omega=-i \tilde g^{-1}d\tilde g=
P_a~\tilde e^a+\frac12 Z_{ab}~\tilde\omega^{ab}+\frac12
M_{ab}~\tilde l^{ab}, \label{MCform2}\ee
where the only independent differentials are $dx^a$.
The one-form $\tilde\Omega$ can be obtained from $\Omega$ in \bref{Omegabb}
by taking the pullback with respect to $x^a\to\T^a$, then
\be d\theta^{a\star}=\frac{\partial\theta^{a}(x)}{\partial
x^b} dx^b ; \ee in such a way every form is defined on $x^a$-space.
One can employ further the one-forms \bref{Omegabb} with $k=0,
b\neq 0$, explicitly calculated  in Sect.5.1.
>From \bref{totalLb} we obtain
\be \tilde e^a=L_P^{a\star}=\left({e^a}_b(x,\theta(x)) + {f^a}_c(\theta(x))
\frac{\partial\theta^{c}(x)}{\partial x^b}\right) dx^b\equiv
{\tilde e^a}{}_b(x,\theta(x)) dx^b. \label{teexb}\ee
where
\bea
{e^a}_b(x,\theta)
&=& {\D^a}_b-b\left(\frac{x^2}{4!}\,{\ep^{a}}_{bc}x^c\,+
\frac12\,((x^c\T_c){\D^a}_b-x^a\T_b) 
\,\right)
\nn\\&+&b^2\,\left(
\left(\frac{x^2\,\T^2}{2!3!}+\frac{(x^2)^3}{7!}\right) {\D^a}_b-
\left(\frac{\T^2}{2!3!}+\frac{(x^2)^2}{7!}\right) {x^a}x_b
-\frac{x^2}{2!3!}\T^a\T_b+\frac{(x\T)}{2!3!}\T^a x_b\right.
\nn\\&+&\left.
\left(\frac{\T^2}{3!} + \frac{(x^2)^2}{5!}\right)\,{\ep^{a}}_{bc}\T^c\,-
\frac{(x^2)}{5!} \,  {\ep^{a}}_{cd}\T^d\,  {x^c}x_b
\,\right)+\CO(b^3),
\label{eexb}\eea
and
\bea
{f^a}_c(\theta(x))&=&
-\,\frac{b}{2}\,{\ep^{a}}_{cb}\T^b+\CO(b^3).
\label{fab}\eea
One can check that  ${f^a}_c(\theta)$
depends only  on $\T^a$  and the dreibein
${\tilde e^a}{}_b$ is linear in the derivatives of the Goldstone fields.

{
In order to construct the action which is invariant under the
$b$-deformed Maxwell group one can use the Volkov-Akulov formula for
invariant \Deq21 action, 
\be
S=\int\,(-\frac1{3!})\ep_{abc}\,L_P^{a\star}L_P^{b\star}L_P^{c\star}=
\int\,d^3 x\, \CL_\theta,\qquad \CL_\theta= \det({\tilde e^a}{}_b).
\label{vaaction}\ee
Using \bref{teexb}-\bref{fab} one can write explicitly  the terms up to $b^2$,
\bea
{\7e^a}{}_b(x,\theta)&=&  {e^a}{}_b(x,\theta)+b\,{h^a}{}_b,\qquad
{h^a}{}_b=- \frac{1}{2}\,{\ep^{a}}_{cd}\T^d \,\frac{\partial\theta^{c}(x)}{\partial x^b}
.
\eea Using
\bea
\det ({\tilde e^a}{}_b)&=&-\frac1{3!}\ep_{abc}\ep^{def}\,{\tilde e^a}{}_d\,{\tilde e^b}{}_e\,{\tilde e^c}{}_f
\nn\\&=&\det ({ e^a}{}_b)\,\left(1+b\,{(e^{-1})^b}_a\,{h^a}{}_b+\frac{b^2}2\,
{(e^{-1})^a}_c\,{(e^{-1})^b}_d\,{h^c}{}_{[a}{h^d}{}_{b]}\right)+\CO(b^3),
\nn\\
\det ({e^a}{}_b)&=&1-b\,
(x\T)+b^2\left(
-\frac{3(x^2)^3}{2240}+\frac{x^2\,\T^2}{12}+\frac{(x\T)^2}3\right)+\CO(b^3),
\nn\\
{(e^{-1})^b}_a&=&{\D^a}_b+b\left(\frac{x^2}{4!}\,{\ep^{a}}_{bc}x^c\,+
\frac12\,((x^c\T_c){\D^a}_b-x^a\T_b)
\,\right)+\CO(b^2)
\eea
we obtain
\bea \CL_\theta&=&\det ({\tilde e^a}{}_b)=1-b\left(
(x\T)+\frac12\ep_{abc}\T^a\frac{\partial\theta^{c}}{\partial x_b}\right)
+b^2\left(
(-\frac{3(x^2)^3}{2240}+\frac{x^2\,\T^2}{12}+\frac{(x\T)^2}3)\right.
\nn\\&-&\left.
\frac{x^2}{48}((x\T){\D_j}^i-x_j\T^i)\frac{\partial\theta^{j}}{\partial x^i}
+\frac{(x\T)}4\ep_{abc}\T^a\frac{\partial\theta^{c}}
{\partial x_b}+
\frac18\,\ep_{abc}\ep^{def}\T^a\T_d \frac{\partial\theta^{b}}{\partial x^e}\frac{\partial\theta^{c}}
{\partial x^f}\right)+\CO(b^3).
\nn\\ \label{VAlag}\eea
{
The lagrangian density \bref{VAlag} contains as one of two terms linear
in $b$ the exact topological lagrangian for \Deq21 Chern-Simons field
\bea \CL^{CS}&=&-\frac{b}2\,{\epsilon_{abc}}\theta^a \frac{\partial\theta^{b}}
{\partial x_c}.
\eea
If we consider higher order terms in $b$ they can be treated
as describing new interaction vertices and 
the Nambu-Goldstone field $\T^a(x)$
looses its topological nature.
The appearance of the terms depending explicitly 
on $x^a$ and $\T^a$ in \bref{VAlag} is
related with the curved geometry in the extended space (see \bref {metricbb}).
Although the explicit formula  \bref{VAlag} looks complicated 
the covariance of the action \bref{vaaction} under the deformed Maxwell group
transformations which describe the group of  motions in the curved space
$(x^a,\T^a)$ follows from our construction obtained by using the nonlinear
realization techniques.}

}

\section{Outlook}

In this paper we consider deformations of the Maxwell algebra. The general
mathematical techniques permit us to solve the problem of complete
classification of these deformations. The commuting generators $Z_{ab}$
in \bref{NHm1} are becoming non-abelian in arbitrary dimension $D$
and are promoted to the $\frac{D(D-1)}{2} $ generators of the $so(D-1,1)$
Lorentz algebra.
The particle dynamics in the $\frac{D(D+1)}{2} $ dimensional coset
\bref{cosetg} becomes the theory of point particles moving on AdS
(for $k>0$) or dS   (for $k<0$) group manifolds in external
electromagnetic fields. If we use standard formula \bref{Lag50} for
the particle action in curved space-time
one can show that the particle moves only in the space-time sector
$(x^a,\T^{ab}=0)$ of the extended  space-time $(x^a,\T^{ab})$ with a
non-local Lorentz force.
The supplementary coordinates $\T^{ab}$ generated by $Z_{ab}$, enter only in MC
one-forms and in particular  they will appear in the model only
in the term representing the
electromagnetic coupling. It is a result of the field equations that the
components of  electromagnetic field strength defined in the basis of
momenta one-forms $L^a$ are constant on-shell(see \bref{constF}).

In "exotic" dimension \Deq21  the symmetry corresponding to the two parameter
deformation of  Maxwell algebra is less transparent.
The coset \bref{cosetg} in \Deq21
if $b\neq0$ is neither the group manifold nor even the symmetric coset space.
In order to find the dynamical realization of deformed Maxwell algebra with 
$b\neq0$ in \Deq21 space-time,
in Sect.5 we consider the \Deq21 field theoretical model obtained by the
assumption that the coordinates $x^a$ are primary and the coordinates
$\T^a$ describe the Goldstone field values. We obtained a non-linear
lagrangian for vector Goldstone field containing the bi-linear kinetic
term describing exactly the  \Deq21 CS Abelian action.

We would like to point out some problems which
deserve still further consideration.

1) The Maxwell algebra was obtained as a deformation of the relativistic Poincare
algebra in the presence of constant electromagnetic background.
One can observe that the relation \bref{NHm1} is dual ( in the sense
of Fourier transformation ) to the canonical non-commutativity
of the Minkowski space-time (see e.g.
\cite{Chaichian:2004za}\cite{Aschieri:2005yw})
\be \left[\hat x^\mu,\hat x^\nu\right]=i\theta^{\mu\nu} \ee
which, as it is argued \cite{Doplicher:1994tu},
describes in algebraic approximation of the quantum gravity
background. It could be interesting to study
this parallelism further.

2) Maxwell algebra contains in four dimensions three quadratic and one
quartic Casimirs \cite{Bacry:1970ye}\cite{Schrader:1972zd}
(see \bref{4Casimir}).
In arbitrary dimension $D$ the Casimir $C_2=Z_{ab}Z^{ab}$
can be incorporated in the particle action by means of the following
extension of the action \bref{Lag}
\be \8L=\pi_a\dot x^a+\frac12f_{ab}(\dot\T^{ab}+\frac12x^{[b}\dot
x^{b]}) -\frac{\lam}{2}(\pi^2+m^2)-\frac{\lam'}{2}(f^2+{m'}^2)
\label{Lag552}\ee where $\lam, \lam'$ are Lagrange multipliers. The
action \bref{Lag552} treats symmetrically the dynamics of $x^a$ and
$\T^{ab}$ variables. One obtains the following second order
lagrangian \be \7L=-m\sqrt{-\dot x^2}-m'
\sqrt{-(\dot\T^{ab}+\frac12x^{[b}\dot x^{b]})^2}. \label{Lag622}\ee
Such a model could possibly relate the additional coordinates $\T^{ab}$
with spin-like degrees of freedom.
It should be interesting to consider the model \bref{Lag552} in
detail and further extend it to the deformed Maxwell algebra geometries,
using the result of sect.3.

3) As we already mentioned, the deformation parameter $k$ with the
dimensionality $[L^{-2}]$ can be described by the formula $|k|=\frac1{R^2}$,
and interpreted as the AdS(dS) radius for $k>0(k<0)$. The  parameter $b$,
with the dimensionality
$[L^{-3}]$, if $k=0$ is related with the closure of the quadrilinear
relation for the following non Abelian translation generators
$P_a$,
 \be
[[P_a,P_b],[P_c,P_d]]=ib\,(\h_{a[c}\ep_{bd]e}-\h_{b[c}\ep_{ad]e})
P^e. \label{PPgeneric1k}\ee
{This relation is an example of higher order
Lie algebra for $n=4$ \cite{Hanlon:1995,DeAzcarraga:1996ts}.
It is an interesting task to understand the
translations  \bref{PPgeneric1k} as describing some \Deq21 dimensional
curved manifold.}

  4) Recently in \cite{Bonanos:2008ez}\cite{Bonanos:2008kr} there
were considered an infinite sequential extensions of the Maxwell
algebra with additional tensorial generators. The concrete form of
these extensions can be determined by studying the
Chevalley-Eilenberg cohomologies at degree two.
The point particle models related with these Poincare algebra
extensions have been studied in \cite{Bonanos:2008kr}.
There appears an interesting question of the dynamical and physical
interpretation of the additional tensorial degrees of
freedom.

{\bf Acknowledgements}

We thank Jorge Alfaro, Sotirios Bonanos, Roberto Casalbuoni, Jaume
Garriga, Gary Gibbons, Mikhail Vasilev, Dimitri Sorokin for
discussions.
 JL would like to thank Universitat de Barcelona for warm hospitality and
acknowledge the support by Polish Ministry of Science and
High Education grant NN202 318534. This work has been partially
supported by MCYT FPA 2007-66665, CIRIT GC 2005SGR-00564, Spanish
Consolider-Ingenio 2010 Programme CPAN (CSD2007-00042).
 J.G. would like to
thank the Galileo Galilei Institute for Theoretical Physics for its
hospitality and INFN for partial support during part of the
elaboration of this work.

\appendix
\section{ Determinatinon of transformation matrix $U(b,k)$}
\subsection{ $U^-(b,k)$ for $\det g<0$}
Here we discuss how the transformation matrix $U^-(b,k)$ in
\bref{transUm} is determined.
We will see $(P,M,Z)$ is related only to dS generators  $(\CP,\CM,\CJ)$
using {\it real component} matrix $U^-(b,k)$
for any $(b,k)$ in the $\det g<0$ region (IV). We fix it becomes
that of $b$-deformation for $k=0$ in \bref{transbdef},
\be
U^-(b,k=0)=U_b.
\ee
Near the $b$-axis we can find $U^-(b,k)$ as the perturbation for small $k$.
It tells a structure of the matrix as
\bea
U^-(b,k)&=&\pmatrix{
\frac{f_3({\kx}) + \frac{{\kx}^2}{9}f_4({\kx})}{\sqrt3b^{1/3}}&
\frac{2{\kx}}{{3\sqrt3}}(f_1({\kx}) + \frac{{\kx}}{3}f_2({\kx}))&
\frac{f_1({\kx}) + \frac{{\kx}}{3}f_2({\kx})}{\sqrt3b^{{2}/{3}}}\cr
-\frac{f_3({\kx}) - \frac{{\kx}^2}{9}f_4({\kx})}{3b^{1/3}}&
\frac23\left(1+\frac{{\kx}}{3}(f_1({\kx}) - \frac{{\kx}}{3}f_2({\kx}))\right)&
\frac{f_1({\kx}) - \frac{{\kx}}{3}f_2({\kx})}{3b^{{2}/{3}}}\cr
\frac{f_3({\kx}) - \frac{{\kx}^2}{9}f_4({\kx})}{3b^{1/3}}&
\frac13\left(1-\frac{2{\kx}}{3}(f_1({\kx}) - \frac{{\kx}}{3}f_2({\kx}))\right)&
-\frac{f_1({\kx}) - \frac{{\kx}}{3}f_2({\kx})}{3b^{{2}/{3}}}\cr
}
\label{Uminus}\eea
where
\be
{\kx}=\frac{k}{b^{{2}/{3}}},\quad 1-\frac{4{\kx}^3}{27}>0,\quad
{\rm for}\quad \det g<0.
\ee
For small ${\kx}$, $f_j({\kx})$'s are polynomials of ${\kx}^3$ and $f_j(0)=1,$
\bea
f_1({\kx})&=&
(1  + {\frac{5\,{{\kx}^3}}{3^4}} + {\frac{44\,{{\kx}^6}}{3^8}} + ...),
\qquad f_2({\kx})=(1  +
  {\frac{7\,{{\kx}^3}}{3^4}}  +
  {\frac{65\,{{\kx}^6}}{3^8}}+...)
\nn\\
f_{3}({\kx})&=&(1  + {\frac{4\,{{\kx}^3}}{3^4}}  + {\frac{35\,{{\kx}^6}}{3^8}} +...),\qquad
f_4({\kx})=(1 +
  {\frac{8\,{{\kx}^3}}{3^4}} +
  {\frac{77\,{{\kx}^6}}{3^8}}+...).
\label{purtxex}\eea
General forms of  $f_j({\kx})$'s are found by requiring that the
$(\CP,\CM,\CJ)$ satisfy the dS algebra at point $(b,k)$ in $\det g<0$.
First from $[\CJ_a,\CJ_b]=-(-i)\ep_{abc}\CJ^c$ we have
\bea
{f_3}({\kx})&=&
\frac{({\7f_1}({\kx}))^2}
{1-\frac23\,{\kx}\,{\7f_1}({\kx})}
+\frac{{\kx}^2}9\,f_4({\kx}),\qquad  \7f_1({\kx})
\equiv{f_1}({\kx})-\frac{{\kx}}3 {f_2}({\kx}).
   \eea
and $\7f_1({\kx})$ satisfies a third order equation
\bea
(1-\frac{4{\kx}^3}{27})({\7f_1}({\kx}))^3
+{\kx}\,{\7f_1}({\kx}) -1=0.
\label{ff12}\eea
Next $[\CM_a,\CM_b]=-(-i)\ep_{abc}\CM^c$ is satisfied using above.
$[\CP_a,\CJ_b]=0$ fix $f_4({\kx})$ as a function of $f_1({\kx})$ and $f_2({\kx})$ as
\bea
{f_4}({\kx})&=&
\frac{3\left(f_2({\kx})(1-\frac{{\kx}}3{\7f_1}({\kx}))-({\7f_1}({\kx}))^2\right)
\;{\7f_1}({\kx})}
{{\kx}\;(1-\frac{2{\kx}}3\,{\7f_1}({\kx}))(1+\frac{{\kx}}3\,{\7f_1}({\kx}))}.
\eea
>From $[\CP_a,\CP_b]=+(-i)\ep_{abc}\CM^c$ we determine $f_2({\kx})$
\bea
{f_2}({\kx})&=&
\frac{1}{{(1-\frac{4{\kx}^3}{27})} \, {f_1}({\kx})}.
\label{ff2}\eea
Using \bref{ff12} and  \bref{ff2} we get
\bea
{{f_1}({\kx})}^3 &=&
\frac{1+\sqrt{1-\frac{4{\kx}^3}{27}}}{2\sqrt{1-\frac{4{\kx}^3}{27}}^3}\;.
\eea
$f_1({\kx})$ is the real cubic root of this equation satisfying $f_1(0)=1$.

In this way we have determined all functions $f_j({\kx})$. They are shown to give
generators verifying all the dS commutation relations in \bref{J1J2alg}.
In small ${\kx}$ expansion they agree with the
perturbative expansion \bref{purtxex} around $b$-deformation ($k=0$).

It is also seen that they are singular on the degenerate curve
\be
-\,A(b,k)=(\frac{b}2)^2-(\frac{k}{3})^3=\frac{b^2}{4}(1-\frac{4{\kx}^3}{27})=+0.
\ee

It is very interesting to see if the generators are analytically continuating
to those of $k^-$-deformation on the $b=0, k<0$ line, that is if
\be
U^-(0,k)=U_k
\ee
holds for $k<0$. It is shown by taking a limit
\be b\to 0,\quad k<0\; ({\rm fixed});\quad
{\kx}=\frac{k}{(b^2)^{\frac13}} \to -\infty.  \ee

In doing it
\be
f_1({\kx})^3\too -\frac{27}{8{\kx}^3}= \frac{27b^2}{-8k^3},\qquad
\ee
then the leading terms of $f_i$'s are
\bea
(f_1,\, f_2,\, f_3, \,f_4)&\to&( \,\frac{3\,b^{2/3}}{-2k}, \,
\frac{9\,b^{4/3}}{2k^2}, \,
\frac{ {\sqrt{3}}\,
        {b^{{{1}/{3}}}}  }{2\,{\sqrt{-{k}}}},\,
{\frac{9\,{\sqrt{3}}\,\,{b^{{{5}/{3}}}}}
    {2\,{\sqrt{-{k}}}^5}}\, ).
 \eea
Taking this limit in \bref{Uminus} it goes to $U_k$ of $k^-$ deformation
in \bref{transkdef}.
\vs

\subsection{ $U^+(b,k)$ for $\det g>0$}

Similarly we determine the transformation matrix $U^+(b,k)$  for $\det g>0$
in \bref{transUp}.
We will see $(P,M,Z)$ is related only to AdS generators  $(\CP,\CM,\CJ)$
using {\it real component} matrix $U^+(b,k)$ for any $(b,k)$ in the
$\det g>0$ region (III). We fix it becomes that of $k^+$-deformation
for $b=0$ in \bref{transkdef},
\be
U^+(b=0,k>0)=U_k.
\ee
Near the $k$-axis we can find $U^+(b,k)$ as the perturbation for small $b$.
It tells a structure of the matrix as
\bea
U^+(b,k)&=&\pmatrix{
\frac1{\sqrt{k}}\,h_3({\bx})&{{\bx}}\,h_4({\bx})&\frac{3{\bx}}{2k}\,h_4({\bx})\cr
-\frac{{\bx}}{\sqrt{k}}\,h_1({\bx})&-{2{\bx}^2}\,h_2({\bx})&-\frac1{k}\,h_5({\bx})\cr
\frac{{\bx}}{\sqrt{k}}\,h_2({\bx})&(1+{2{\bx}^2}\,h_1({\bx}))&\frac1{k}\,(1+{3{\bx}^2}\,h_1({\bx}))}
\label{Uplus}\eea
where
\be
{\bx}=\frac{b}{k^{{3}/{2}}},\quad 1-\frac{27{\bx}^2}{4}>0,\quad
{\rm for}\quad \det g>0.
\ee
For small ${\bx}$, $h_j({\bx})$'s are polynomials of ${\bx}^2$ and $h_j(0)=1,$
\bea
h_1({\bx})&=&1  + {4\,{{\bx}^2}} + {21\,{{\bx}^4}} + ...,\qquad
h_2({\bx})=1  +  {5\,{{\bx}^2}}  + {28\,{{\bx}^4}}+...
\nn\\
h_{3}({\bx})&=& 1  + {\frac{15\,{{\bx}^2}}{2^3}}  + {\frac{1155\,{{\bx}^4}}{2^7}} +...,\quad
h_4({\bx})=1 +  {\frac{35\,{{\bx}^2}}{2^3}} + {\frac{3003\,{{\bx}^4}}{2^7}}+...,\quad
\nn\\
h_5({\bx})&=&1 +  {3\,{{\bx}^2}} + {15\,{{\bx}^4}}+....
\label{purtxexk}\eea
General forms of  $h_j({\bx})$'s are found by requiring that the
$(\CP,\CM,\CJ)$ satisfy the AdS algebra \bref{D3dMax3} with $k>0$
at point $(b,k)$ in $\det g>0$.

The results are
\bea
{{h_1}({\bx})}&=&
    {{\frac{ {h_2}({\bx})\,
         \left( 1 + 2\,{{\bx}^2}\, {h_2}({\bx}) \right) }{1 +
         3\,{{\bx}^2}\, {h_2}({\bx})}}},\nn\\
{ {h_3}({\bx})}&=&
   {{\frac{ \,
        \left( 2 + 3\,{{\bx}^2}\, {h_2}({\bx}) \right) \,
        }{2\,
        \left( 1 + 2\,{{\bx}^2}\, {h_2}({\bx}) \right) }}}\sqrt{
   {{\frac{ \left( 1 + 3\,{{\bx}^2}\, {h_2}({\bx}) \right)}
      {\left( 1 - {\frac{27\,{{\bx}^2}}{4}} \right) \,{h_2}({\bx})\,
         }}}} ,\nn\\
{{{ {h_4}({\bx})}}}&=&\sqrt{
   {{\frac{ {h_2}({\bx})}
      {\left( 1 - {\frac{27\,{{\bx}^2}}{4}} \right) \,
        \left( 1 + 3\,{{\bx}^2}\, {h_2}({\bx}) \right) }}}},\nn\\
{ {h_5}({\bx})}&=&
   {{\frac{1 + {\frac{9\,{{\bx}^2}\, {h_2}({\bx})}{4}}}
      {\left( 1 - {\frac{27\,{{\bx}^2}}{4}} \right) \,
        {{\left( 1 + 3\,{{\bx}^2}\, {h_2}({\bx}) \right) }^2}}}},\eea
and $h_2({\bx})$ is determined by a third order equation
\bea
1 - \left( 1 - 9\,{{\bx}^2} \right) \,{h_2}({\bx}) -
  4\,{{\bx}^2}\,\left( 1 - {\frac{27\,{{\bx}^2}}{4}} \right) \,
   {{{h_2}({\bx})}^2} -
  4\,{{\bx}^4}\,\left( 1 - {\frac{27\,{{\bx}^2}}{4}} \right) \,
   {{{h_2}({\bx})}^3}=0,
\eea
whose real solution is
\bea
{ {h_2}({\bx})}&=&\frac1{{3\,{{\bx}^2}}}\left( \frac1{\sqrt{1 - {\frac{27\,{{\bx}^2}}{4}}}}
\,\cos \left(\frac13{\arctan
({\frac{3\,{\sqrt{3}}\,{\bx}}
               {2\,{\sqrt{1 - {\frac{27\,{{\bx}^2}}{4}}}}}})}\right)-1\right).
\eea

In this way we have determined all functions $h_j({\bx})$
having the small ${\bx}$ expansion  in \bref{purtxexk} thus
$U^+(b,k)$ becomes $U_k$ for $b=0,k>0$.
It is also seen that they are singular on the degenerate curve
\be
\,A(b,k)=(\frac{k}{3})^3-(\frac{b}2)^2=(\frac{k}{3})^3(1-\frac{27{\bx}^2}{4})=+0.
\ee

\vs
\section{ $f_j$ and $F_j $}
Here we give results of summations of functions $f_j$ in \bref{funcfi}
and $F_j $ in \bref{funcFi}.  $f_j(Y)\equiv f_j(\A=1,Y)$ are\footnote{
For $x^a$ is timelike
$Y^6=-b^2r^6,\,r=\sqrt{-x^2}.$
In this case trigometric functions and trigonometric ones are interchanged. }
\bea
f_0(\A,Y)&=&
 {\frac{1}{3}}\left(2\,\cos ({\frac{{\sqrt{3}}\,{\A}\,Y}{2}})\,
       \cosh ({\frac{{\A}\,Y}{2}}) + \,\cosh ({\A}\,Y)\right),
\nn\\
f_1(\A,Y)
&=&
  \frac{1}{3Y}\left( {\sqrt{3}}\,
          \sin ({\frac{{\sqrt{3}}\,{\A}\,Y}{2}})\cosh ({\frac{{\A}\,Y}{2}})\, +
         \cos ({\frac{{\sqrt{3}}\,{\A}\,Y}{2}})\,
          \sinh ({\frac{{\A}\,Y}{2}})   +
      \sinh ({\A}\,Y)\right),
\nn\\
f_2(\A,Y)
&=&
  \frac{1}{3\,{Y^2}}\left( {\sqrt{3}}\,
       \sin ({\frac{{\sqrt{3}}\,{\A}\,Y}{2}})\,\sinh ({\frac{{\A}\,Y}{2}})\,
       - \cos ({\frac{{\sqrt{3}}\,{\A}\,Y}{2}})\,
         \cosh ({\frac{{\A}\,Y}{2}})  +
      \cosh ({\A}\,Y) \right),
\nn\\
f_3(\A,Y)
&=& \frac{1}{3\,{Y^3}}\left(
  -2\,\cos ({\frac{{\sqrt{3}}\,{\A}\,Y}{2}})\,
       \sinh ({\frac{{\A}\,Y}{2}}) + \,\sinh ({\A}\,Y)\right),
\nn\\
f_4(\A,Y)
&=&\frac{1}{3\,{Y^4}}\left(
-
      {\sqrt{3}}\,\sin ({\frac{{\sqrt{3}}\,{\A}\,Y}{2}})\,
       \sinh ({\frac{{\A}\,Y}{2}})- \cos ({\frac{{\sqrt{3}}\,{\A}\,
              Y}{2}})\,\cosh ({\frac{{\A}\,Y}{2}})
       + \cosh ({\A}\,Y) \right) ,
\nn\\
f_5(\A,Y)
&=&\frac{1}{3\,{Y^5}}\left(
  - {\sqrt{3}}\,
         \sin ({\frac{{\sqrt{3}}\,{\A}\,Y}{2}})\, \cosh ({\frac{{\A}\,Y}{2}}) +
      \cos ({\frac{{\sqrt{3}}\,{\A}\,Y}{2}})\,
       \sinh ({\frac{{\A}\,Y}{2}}) + \sinh ({\A}\,Y) \right)
\nn\\ \label{fj}
    \eea
 and $F_j(Y)$'s are
\bea
F_0(Y)&=&\frac{1}{3Y}\left(
 {\sqrt{3}}\,\cosh ({\frac{Y}{2}})\,
    \sin ({\frac{{\sqrt{3}}\,Y}{2}}) +
   \cos ({\frac{{\sqrt{3}}\,Y}{2}})\,
    \sinh ({\frac{Y}{2}}) + \sinh (Y)\right),
\nn\\
F_1(Y)&=&\frac{1}{3Y^2}\left(
  {\sqrt{3}}\,\sin ({\frac{{\sqrt{3}}\,Y}{2}})\,
    \sinh ({\frac{Y}{2}}) - \cos ({\frac{{\sqrt{3}}\,Y}{2}})\,
      \cosh ({\frac{Y}{2}})   + \cosh (Y)
  \right),
\nn\\
F_2(Y)&=&
  \frac{1}{3\,{Y^3}}\left(  -2\,\cos ({\frac{{\sqrt{3}}\,Y}{2}})\,
    \sinh ({\frac{Y}{2}}) + \sinh (Y)\right),
\nn\\
F_3(Y)&=& \frac{1}{3\,{Y^4}}\left( -
   {\sqrt{3}}\,\sin ({\frac{{\sqrt{3}}\,Y}{2}})\,
    \sinh ({\frac{Y}{2}})
  - \cos ({\frac{{\sqrt{3}}\,Y}{2}})\,
      \cosh ({\frac{Y}{2}})   + \cosh (Y)\right),
\nn\\
F_4(Y)&=&\frac{1}{3\,{Y^5}}\left(
  - {\sqrt{3}}\,\cosh ({\frac{Y}{2}})\,
      \sin ({\frac{{\sqrt{3}}\,Y}{2}})   +
   \cos ({\frac{{\sqrt{3}}\,Y}{2}})\,
    \sinh ({\frac{Y}{2}}) + \sinh (Y)\right),
\nn\\
F_5(Y)&=&\frac{1}{3\,{Y^6}}\left(
  -3 + 2\,\cos ({\frac{{\sqrt{3}}\,Y}{2}})\,
    \cosh ({\frac{Y}{2}}) + \cosh (Y)\right)
\eea


For small $b$ we have expansions of $f_j$ and $F_j$ as
\bea
F_i(Y)&=&\frac{1}{(i+1)!}+\frac{b^2\,(x^2)^3}{(i+7)!}+\CO(b^4),\quad
\nn\\
F_i(Y')&=&\frac{1}{(i+1)!}
+\CO(b^4),\qquad
f_i(Y')=\frac{1}{i!}
+\CO(b^4).
\label{funcFibex}\eea
Keeping up to $b^2$,
\bea
\pmatrix{L_{0P}^a\cr L_{0M}^a\cr L_{0Z}^a}&=&
\left[ \pmatrix{1\cr 0\cr 0}{\D^a}{}_c+
\pmatrix{b^2(x^2)^3/7!\cr bx^2/3!\cr-b(x^2)^2/5!}
{O^a}{}_c+\pmatrix{-bx^2/4!\cr-b^2(x^2)^2/6! \cr \frac12+b^2(x^2)^3/8! }
{\ep^{a}}_{cb}x^b \right]\,dx^c.
\nn\\ \label{L0pmzbx}\eea
and
\bea
\pmatrix{L_{P}^a\cr L_{M}^a\cr L_{Z}^a}&=&
\left[ \pmatrix{0\cr 0\cr 1}{\D^a}{}_c+
\pmatrix{0\cr -b^2\T^2/3!\cr 0 }
{\7O^a}{}_c+\pmatrix{-b/2!\cr 0 \cr b^2\T^2/4! }
{\ep^{a}}_{cb}\T^b \right]\,d\T^c.
\nn\\ &+&\left[(I_3)\,{\D^a}{}_c+\pmatrix{V_O}{\7O^a}{}_c+\pmatrix{V_E}
{\ep^{a}}_{cb}\T^b \right]\,
\pmatrix{L_{0P}^c\cr L_{0M}^c\cr L_{0Z}^c},
\eea
where  $L_0^c$'s are given in \bref{L0pmzbx} and
\bea
\pmatrix{V_O}&=&\pmatrix{0&b\T^2/2!& 0\cr
0&0&-b^2\T^2/2!\cr -b\T^2/2!&-b^2(\T^2)^2/4!&0},
\nn\\
\pmatrix{V_E}&=&\pmatrix{b^2\T^2/3!&0&-b\cr
b&b^2\T^2/3!&0\cr 0&-1&b^2\T^2/3!}.
\eea
Then up to $b^2$ we have the result of \bref{MCbexp}.

\end{document}